\pdfoutput=1

\documentclass[11pt]{article}
\usepackage{jheppub}

\usepackage{amsmath}
\usepackage{amssymb}
\usepackage{graphicx,color,slashed,braket}
\usepackage{bbm}
\usepackage{ifpdf}
\usepackage{comment}
\usepackage{soul}

\usepackage{booktabs}
\usepackage{multirow}
\usepackage{makecell}
\usepackage{float}
\usepackage{verbatim}
\usepackage{caption}
\usepackage{cancel}
\usepackage{enumerate}
\usepackage{tcolorbox}
\usepackage[font=small,labelfont=bf]{caption}

\usepackage{color}
\usepackage{colortbl}
\definecolor{darkred}{rgb}{0.7,0,0}
\definecolor{darkgreen}{rgb}{0.1,0.4,0}
\definecolor{darkblue}{rgb}{0.3,0.3,0.7}
\definecolor{lightblue}{rgb}{0.8,0.8,1}

\usepackage[percent]{overpic}
\usepackage{tikz}
\usepackage{graphicx}
\usetikzlibrary{calc}

\usepackage{hyperref}
\usepackage{setspace}
\hypersetup{
	colorlinks=true,
	linkcolor=darkblue,
	citecolor=darkred,
	urlcolor=darkgreen,
	linktoc=page
}
\newcommand{\be}{\begin{equation}}
\newcommand{\ee}{\end{equation}}

\def\be{\begin{equation}}
\def\ee{\end{equation}}
\def\bea{\begin{eqnarray}}
\def\eea{\end{eqnarray}}

\newcommand{\e}{\mathrm{e}}

\newcommand{\dd}{\mathrm{d}}

\newcommand{\Mpl}{M_\text{Pl}}

\newcommand{\vol}{\text{vol}}

\allowdisplaybreaks

\title{Supersymmetric scale-separated AdS\textsubscript{3} orientifold vacua of type IIB}

\preprint{UUITP-04/25}

\author{Vincent Van Hemelryck}

\affiliation{Department of Physics and Astronomy, Uppsala University, Box 516, SE-75120, Uppsala, Sweden}

\emailAdd{vincent.vanhemelryck@physics.uu.se}

\notoc

\abstract{I construct supersymmetric AdS$_3$ vacua of type IIB string theory that exhibit parametric scale separation in the controlled regime. These solutions arise from compactifications on seven-dimensional manifolds equipped with co-closed $G_2$-structures, in the presence of orientifold planes preserving minimal supersymmetry. I focus on $\mathbb{Z}_2^3$ orbifolds of a specific solvmanifold and all seven-dimensional nilmanifolds, each requiring distinct configurations of intersecting O5-planes, which are treated in the smeared approximation. Weak string coupling and large volumes can be achieved for classical backgrounds on both nilmanifolds and solvmanifolds by tuning unbounded fluxes to large values. This achieves scale separation in the nilmanifold case, but the same cannot be concluded for the solvmanifolds. Furthermore, for some solutions, the holographic field theory operators dual to the lightest scalar fields have integer conformal dimensions at tree level, as in other scale-separated type IIA orientifold backgrounds.
}
\begin{document}

\maketitle

\newpage
\tableofcontents
\vspace{0.5cm}
\hrule
\section{Introduction}
\label{sec:Introduction}
Establishing full moduli stabilisation while also realising a large enough hierarchy between the cosmological radius and the Kaluza-Klein radius remains a tremendous challenge in flux compactifications. Nevertheless, this is one of the most standard features that phenomenological compactifications need to exhibit, such that the vacua are truly lower-dimensional. However, most AdS compactifications of string theory do not satisfy the scale separation condition:
\begin{equation}
    \frac{L_\mathrm{KK}}{L_\mathrm{AdS}} \ll 1\,,
\end{equation}
where $L_\mathrm{KK}$ and $L_\mathrm{AdS}$ are the KK and AdS radius respectively.
Some recent progress has provided strong arguments against scale separation in specific settings. For instance, refs.~\cite{Cribiori:2022trc,Bobev:2023dwx,Cribiori:2023ihv,Perlmutter:2024noo} indicate that scale separation for compactifications of several theories with extended supersymmetry is unattainable. Moreover, it has been argued in ref.~\cite{Gautason:2015tig} that for compactifications without sources, scale separation cannot occur for isotropic manifolds of positive scalar curvature, more precisely for when the KK scale decouples from the scalar curvature, and due to the lack of concrete examples it was conjectured not to happen in ref.~\cite{Collins:2022nux}.
All these findings support the strong version of the Anti-de Sitter distance conjecture \cite{Lust:2019zwm}, implying that no supersymmetric scale-separated vacua exist. 
However, the minimally supersymmetric theories with sources are not constrained yet by the arguments of refs.~\cite{Cribiori:2022trc,Bobev:2023dwx,Cribiori:2023ihv,Perlmutter:2024noo,Gautason:2015tig,Collins:2022nux}, and indeed, there are a few scale-separated models in string theory, therefore violating the Anti-de Sitter distance conjecture \cite{Lust:2019zwm}. Nevertheless, their validity is questioned and is an active research topic in the context of the swampland program. For a recent review on this issue, see ref.~\cite{Coudarchet:2023mfs}.

The current state of the art of supersymmetric scale-separated vacua can be summarised as a type IIA vs. IIB dichotomy: in type IIA, there are scale-separated vacua where all moduli are stabilised using fluxes and orientifold planes, hence they are referred to as classical vacua. The separation of scales is achieved by tuning unbounded, integer-quantised fluxes to infinity. Examples in four dimensions are compactifications on manifolds with SU(3)-structure, usually toroidal orbifolds, e.g. the DGKT solutions in massive IIA \cite{DeWolfe:2005uu,Camara:2005dc} and analogue solutions in massless IIA \cite{Cribiori:2021djm,Carrasco:2023hta}, of which some can be lifted to solutions of M-theory which appear to have weak $G_2$-structure \cite{VanHemelryck:2024bas}.
Examples in three dimensions are compactifications on manifolds with $G_2$-holonomy, as considered in \cite{Farakos:2020phe,VanHemelryck:2022ynr}.\footnote{See also refs.~\cite{Farakos:2023nms,Farakos:2023wps,Tringas:2023vzn} for variations on these setups, and ref.~\cite{Andriot:2025cyi} for a discussion on scale separation in rolling solutions.}
Although the holographic dual field theories for these models are not known, refs.~\cite{Conlon:2021cjk, Apers:2022tfm,Quirant:2022fpn} found, quite surprisingly, that the mass spectrum of the scalars in the four-dimensional scale-separated vacua is such that the putative holographic dual operators have integer conformal dimensions.\footnote{This feature is expected to break down by small one-loop corrections, see ref.  \cite{Plauschinn:2022ztd}.} This feature does not appear in the three-dimensional vacua of ref.~\cite{Farakos:2020phe}, as shown in ref.~\cite{Apers:2022zjx}.

However, these constructions have received criticism from several angles. For instance, these models all contain O6-planes that intersect in the covering space of the orbifold, and their charge density has been smeared over compact space, i.e. they are not treated as the codimension-3 objects that they are. However, this criticism has been refuted for a large part as the local orientifold backreaction has been computed and shown to be small \cite{Baines:2020dmu,Junghans:2020acz,Marchesano:2020qvg, Andriot:2023fss,Emelin:2024vug}. Moreover, the O6-planes need not intersect once the orbifold singularities are blown up \cite{Junghans:2023yue}. Nevertheless, the massive type IIA models cannot be lifted to M-theory due to the non-zero Romans mass. 
The models in massless type IIA \cite{Cribiori:2021djm,Carrasco:2023hta} overcome some of these issues, as they do have the potential to lift to M-theory on manifolds with weak $G_2$-structure, as recently shown in ref.~\cite{VanHemelryck:2024bas} and where the O6-planes geometrise.
Additionally, it has been argued recently in ref.~\cite{Montero:2024qtz} that the massive IIA backgrounds are in tension with the weak gravity conjecture for branes \cite{Arkani-Hamed:2006emk}, as the moduli space of the 3d theory living on a D4-domain wall is lifted due to the absence of a parity symmetry on the brane worldvolume. The massless type IIA vacua do not suffer from this problem.

The situation in type IIB is quite different. Non-perturbative effects are usually necessary for full moduli stabilisation and obtaining scale separation. Examples are four-dimensional Calabi-Yau compactifications with three-form fluxes and a gaugino condensate or Euclidean D3-instantons, such as in the KKLT and LVS scenarios \cite{Kachru:2003aw,Conlon:2005ki,Balasubramanian:2005zx}. Recent progress has been made in finding solutions with a large hierarchy of scales (i.e. finding a small flux superpotential), see for example refs.~\cite{Demirtas:2019sip,Demirtas:2020ffz,Demirtas:2021nlu,Demirtas:2021ote,McAllister:2024lnt}, however, these scenarios have been criticised on different fronts as well, see e.g. refs.~\cite{Bena:2018fqc,Gao:2020xqh,Bena:2020xrh, Gao:2022fdi,Gao:2022uop,Lust:2022lfc} for the most recent criticism. So far, no supersymmetric and scale-separated vacua of type IIB have been found where only classical ingredients stabilise all moduli, while maintaining weak coupling and large volume.\footnote{The 4d vacua of refs.~\cite{Caviezel:2009tu,Petrini:2013ika} are not well-controlled, as some one-cycles are becoming small in string units, see ref.~\cite{Cribiori:2021djm}. Recently, the possibility of scale separation in type IIB orientifold compactifications on flat manifolds has been studied in \cite{Tringas:2025uyg}.}

However, this paper presents the first such examples. These are realised by compactifying type IIB string theory on seven-dimensional manifolds equipped with a co-closed $G_2$-structure. Compactifications on such backgrounds have been considered before in refs.~\cite{Passias:2020ubv, Emelin:2021gzx} and \cite{VanHemelryck:2022ynr}. 
In particular, this paper studies $\mathbb{Z}_2^3$ orbifolds of two main classes of group manifolds: solvmanifolds and nilmanifolds. For the former, a specific solvmanifold is considered, which has been described in the literature before, see e.g. refs.~\cite{DallAgata:2005zlf,Andriot:2015sia, Manero2020315}. After orientifolding these specific spaces with an O5-involution, one obtains one class of manifolds with four (intersecting) sets of (smeared) O5-planes and another class with only one set of O5-planes. The former is reminiscent of the type IIA vacua with 4 sets of (intersecting) O6-planes, whereas the latter case can be regarded as the supersymmetric cousin of a scale-separated, but non-supersymmetric type IIB vacuum found in ref.~\cite{Arboleya:2024vnp,Arboleya:2025ocb}, which are related through skew-whiffing. These solutions achieve large cycle volumes, weak coupling and scale-separation between the AdS scale and cycle volumes, but those do not seem to provide a good estimate of the KK scale.
The second class of manifolds comprises nilmanifolds, for which the latter issue does not arise. The study of seven-dimensional nilmanifolds that admit a co-closed $G_2$-structure has received great attention recently, e.g. in refs.~\cite{MR2811660,MR3739330,MR4626831,MR4789076}. There are only 12 nilmanifolds consistent with the orbifold $\mathbb{Z}_3^2$, and scale-separated solutions in the controlled regime were found in this paper on only 6 of them. The simplest nilmanifold, being a direct product of the 3d Heisenberg nilmanifold and a flat four-torus, exhibits integer conformal dimensions for the operators dual to the scalars, all the others do not. 

The analysis here uses the tools of refs.~\cite{Emelin:2021gzx,Passias:2020ubv,VanHemelryck:2022ynr}, but is different in the sense that in those references, models with seven (intersecting) sets of O5-planes were considered instead of fewer. Particularly in ref.~\cite{Emelin:2021gzx}, it was argued that scale separation can be achieved in their setup if one includes additional D5-branes to cancel some of the tadpoles, and if one can make the curvature small compared to the compactification radius. However, the latter clashes with geometric flux quantisation when the manifold is isotropic.
The setups of this paper circumvent these issues in the sense that the manifolds are non-isotropic, and one also does not necessarily need wrapped D5-branes due to the choice of orbifold generators.

This paper is organised as follows: section~\ref{sec:orbifolds} discusses the $\mathbb{Z}_2^3$ orbifold and orientifold of seven-dimensional twisted tori and their relation to co-closed $G_2$-structures. Section~\ref{sec:G2-solutions} briefly reviews type IIB compactifications on manifolds with $G_2$-structure, and discusses the specific compactifications on some solvmanifolds and nilmanifolds. Finally, the results are summarised in section~\ref{sec:summary}.

\section{\texorpdfstring{$\mathbb{Z}_2^3$}{Z2 x Z2 x Z2} orbifolds of twisted tori and co-closed \texorpdfstring{$G_2$}{G2}-structures}
\label{sec:orbifolds}
Before discussing three-dimensional flux compactifications of type IIB string theory, it is useful to focus on the geometries and especially the orbifolds that are used for this purpose.
Especially nil- and solvmanifolds play a special role here, which are group manifolds arising from a nilpotent or solvable group, respectively.
They can be regarded as a twisted seven-torus, denoted by $\tilde{\mathbb{T}}^7$, for which the metric in the vielbein parametrisation is
\begin{equation}
\label{eq:metric}
    \dd s_7^2 = \sum_{i=1}^7 L_i^2 (e^i)^2\,,
\end{equation}
where the $e^i = e^i_m \dd x^m$ are the unit length local coframe one-forms and the $L_i$ are the associated radii, expressed in string units ($2\pi \ell_s =1)$. These one-forms are not necessarily closed, and indeed, they satisfy the Maurer-Cartan equations:
\begin{equation}
\label{eq:MaurerCartan}
    \dd e^a = \frac{1}{2}f^{a}{}_{{}{bc}} \;e^b \wedge e^c\,,
\end{equation}
where the $f^{a}{}_{{}{bc}}$ are the structure constants of the Lie algebra underlying the group manifold.
To project out certain forms of the seven-dimensional manifold and to generate multiple orientifold plane images, it is useful to consider orbifolds of this space. In this work, the $\Gamma=\mathbb{Z}_2^3$ orbifold is considered, very similar to refs.~\cite{DallAgata:2005zlf,Emelin:2021gzx}.
The three $\mathbb{Z}_2$ generators $\alpha$, $\beta$ and $\gamma$ act on the coframe one-forms as follows:
\begin{align}
\label{eq:OrbifoldGammaEs}
\begin{split}
    &\alpha(e) = (-e^1, -e^2, e^3, e^4, -e^5, -e^6, e^7)\,,\\
    &\beta(e) = (e^1, e^2, -e^3, -e^4, -e^5, -e^6, e^7)\,,\\
    &\gamma(e) = (-e^1, e^2, -e^3, e^4, -e^5, e^6, -e^7)\,.
\end{split}
\end{align}
The Maurer-Cartan equations must be invariant under the orbifold actions, which do not allow for all structure constants, leaving only the following ones:
\begin{equation}
\label{eq:allowed_structure_constants}
    f^{5}{}_{14}, \quad f^{5}{}_{23}, \quad f^{6}{}_{13}, \quad f^{6}{}_{24}, \quad f^{7}{}_{12}, \quad f^{7}{}_{34}, \quad f^{7}{}_{56} \quad +\; \text{permutations}\,.
\end{equation}
To investigate the number of fixed points of the orbifold and orientifold later, one needs to work in a coordinate representation of the metric and coframe one-forms, with coordinates $x^1,...,x^7$. 
It is instructive to consider the standard $\Gamma=\mathbb{Z}_2^3$ orbifold of the \textit{ordinary} torus $\mathbb{T}^7$ first, as it will act similarly on the coordinate systems of the solv- and nilmanifolds. The elements of $\Gamma$ can be written as follows:
\begin{align}
\label{eq:OrbifoldGammaXS}
\begin{split}
    &\alpha(x) = (-x^1, -x^2, x^3, x^4, -x^5, -x^6, x^7)\,,\\
    &\beta(x) = (x^1, x^2, -x^3, -x^4, -x^5, -x^6, x^7)\,,\\
    &\gamma(x) = (-x^1, x^2, -x^3, x^4, -x^5, x^6, -x^7)\,.
\end{split}
\end{align}
These three $\mathbb{Z}_2$ actions square to one and each have 16 fixed points, just as their combinations $\alpha\beta$, $\gamma\beta$, $\alpha\gamma$, $\alpha\beta\gamma$. For example, the fixed points of $\gamma(x) = x $ are at $x^{1,3,5,7}=\{0, \frac{1}{2}\}$, for all $x^{2,4,6}$.
After orbifolding, one should orientifold the space by an O5-involution for further purposes of this paper, similar to ref.~\cite{Emelin:2021gzx}. Such an involution that also leaves the Maurer-Cartan equations \eqref{eq:MaurerCartan} invariant, is for instance:
\begin{equation}
\label{eq:O5_involution}
    \sigma_\gamma(x) = (-x^1, x^2, -x^3, x^4, -x^5, x^6, -x^7)\,,
\end{equation}
acting on the coordinates in the same way as $\gamma$. The hypersurfaces of fixed points of this orientifold involution correspond to O5-planes, as they are of codimension 4.
Since the involution acts the same as $\gamma$, the orientifold $\sigma_\gamma$ has the same fixed points as  $\gamma$, at $x^{1,3,5,7}=\{0, \frac{1}{2}\}$. Moreover, all orbifold images of $\sigma_\gamma$, i.e. $\sigma_\gamma \alpha$, $\sigma_\gamma {\alpha\beta}$, ... have fixed points too, which all lead to seven different (intersecting) sets of O5-planes. Finally, $\sigma_\gamma \gamma$ leaves everything invariant and leads to an O9-plane, meaning that one can consider type I string theory as well, after including 32 D9-branes.
This orbifold and orientifold were already considered in ref.~\cite{Emelin:2021gzx}, and the inclusion of the O9-plane was useful for matching results with 3d compactifications of heterotic string theory, see e.g. \cite{deIaOssa:2019cci,delaOssa:2021cgd,delaOssa:2024dzo}, due to heterotic - type I string theory duality.
Nevertheless, it is not well-understood how the orbifold singularities of such spaces can be resolved.

For this work, it is important to reduce the number of intersecting O5-plane sets and additionally remove the O9-plane so that one remains within type IIB string theory. This is achieved by including shifts in the orbifold actions as well. Such types of orbifolds were crucial in Joyce's construction of compact spaces with $G_2$-holonomy \cite{Joyce:1996i,Joyce:1996ii}. This can be achieved by adding 7-vectors $\Vec{a}$, $\Vec{b}$ and $\Vec{c}$ whose entries are either 0 or $1/2$:
\begin{align}
\label{eq:OrbifoldGammaHat}
\begin{split}
    &\hat \alpha(x) = \biggl(-x^1, -x^2, x^3, x^4, -x^5, -x^6, x^7 \biggr)+\Vec{a}\,,\\
    &\hat \beta(x) = \biggl(x^1, x^2, -x^3, -x^4, -x^5, -x^6, x^7\biggr)+\Vec{b}\,,\\
    &\hat \gamma(x) = \biggl(-x^1, x^2, -x^3, x^4, -x^5, x^6, -x^7\biggr)+\Vec{c}\,.
\end{split}
\end{align}
As a first example, which is relevant for the solvmanifold, one should consider the orbifold $b_7 = c_1 = c_3 = c_5 = 1/2$ and all the rest are vanishing.
\begin{equation}
\label{eq:orbifold_Gamma4}
    \hat{\Gamma}_4\;: \qquad  b_7 = c_1 = c_3 = c_5 = 1/2\,.
\end{equation}
For this orbifold, only the elements $\hat \alpha$ and $\hat \gamma$ have fixed points (16 each), all the other combinations do not. 
According to the criteria of Joyce \cite{Joyce:1996i,Joyce:1996ii}, this orbifold on the \textit{ordinary} $\mathbb{T}^7$ has singularities of the `good' type, meaning that they do not intersect and that a blow-up procedure can resolve them. Every fixed point is locally a three-torus on the \textit{ordinary} $\mathbb{T}^7$. The group element $\hat{\gamma}$ reduces the set of independent fixed $\mathbb{T}^3$'s of $\hat{\alpha}$ by half, to 8, hence there are eight singularities of the local form $\mathbb{R}^4/\mathbb{Z}_2\times \mathbb{T}^3$. Similarly, the fixed points of $\hat{\gamma}$ are shuffled around by both $\hat{\alpha}$ and $\hat{\beta}$, leading to only 4 independent singularities. The blow-up procedure generates three local three-cycles for every independent singularity, leading to $(8+4)\times 3 = 36$ additional three-cycles, whose associated blow-up moduli are said to belong to the \textit{twisted} sector. There are also 12 local two-cycles from the blow-up. 

Further orientifolding with the same involution $\sigma_\gamma$ of eq.~\eqref{eq:O5_involution} results in fixed points of codimension four for $\sigma_\gamma$, $\sigma_\gamma \hat \alpha$, $\sigma\hat \beta$ and $\sigma_\gamma {\hat \alpha \hat  \beta}$ , leading to 4 sets of O5-planes. These are summarised in Table~\ref{tab:O5sGammaHat4}. The actions $\sigma_\gamma \hat\gamma$, $\sigma_\gamma{\hat\alpha \hat\gamma}$, $\sigma_\gamma{\hat\beta \hat\gamma}$, $\sigma_\gamma {\hat\alpha \hat\beta \hat\gamma}$ have no fixed points: for each of them, there is at least one coordinate $x^i$ that is mapped into $x^i +1/2$, hence no further orientifold images are created, neither does the orientifold involution reduce the amount of independent orbifold fixed points. The orientifold planes are summarised in Table~\ref{tab:O5sGammaHat4}.
\begin{table}[t]
    \centering
    \begin{tabular}{|l||c|c|c|c|c|c|c|}\hline
      {$\hat{\Gamma}_4$} & $x^1$ & $x^2$ & $x^3$ & $x^4$ & $x^5$ & $x^6$ & $x^7$  \\ \hline \hline 
        O5$_{\sigma_\gamma}$ & $-$ & $\times$ & $-$ & $\times$ & $-$ & $\times$ & $-$\\ \hline
       O5$_{\sigma_\gamma\hat\alpha}$ & $\times$ & $-$ & $-$ & $\times$ & $\times$ & $-$ & $-$\\ \hline
       O5$_{\sigma_\gamma\hat\beta}$ & $-$ & $\times$ & $\times$ & $-$ & $\times$ & $-$ & $-^*$\\ \hline
       O5$_{\sigma_\gamma\hat\alpha\hat\beta}$ & $\times$ & $-$ & $\times$ & $-$ & $-$ & $\times$ & $-^*$\\ \hline
    \end{tabular}
    \caption{The different orientifold planes after orientifolding the orbifold $\hat \Gamma_4$. The crosses indicate the directions parallel to the O5-plane, while the dashes correspond to the directions in which the O5-planes are localised, at the values $\{0,1/2\}$ and $\{1/4, 3/4\}$, the latter being indicated with an asterisk.}
    \label{tab:O5sGammaHat4}
\end{table}

Another example arises from introducing more shifts, for instance, by choosing 
\begin{equation}
\label{eq:orbifold_Gamma1}
   \hat{\Gamma}_1\;: \qquad a_2 = a_4 = b_2 = b_6 = c_1 = c_2 = c_3 = 1/2\,.
\end{equation}
An orbifold like this appeared in ref.~\cite{Andriolo:2018yrz} but without the shifts in $\hat{\gamma}$. Only the orbifold action $\hat{\gamma}$ has fixed points, and again, Joyce's criterion guarantees that this orbifold of the \textit{ordinary} torus has singularities which can be blown up, of which there are only 4. 
Furthermore, only the orientifold involution $\sigma_\gamma$ has fixed points, all orbifold images do not have any. Hence, there is only one set of O5-planes, summarised in Table~\ref{tab:O5sGammaHat1}. The orientifold involution does not further identify different orbifold singularities here either.
\begin{table}[ht]
    \centering
    \begin{tabular}{|l||c|c|c|c|c|c|c|}\hline
      {$\hat{\Gamma}_1$} & $x^1$ & $x^2$ & $x^3$ & $x^4$ & $x^5$ & $x^6$ & $x^7$  \\ \hline \hline 
        O5$_{\sigma_\gamma}$ & $-$ & $\times$ & $-$ & $\times$ & $-$ & $\times$ & $-$\\ \hline
    \end{tabular}
    \caption{The different orientifold planes after orientifolding the orbifold $\hat \Gamma_1$. The crosses indicate the directions parallel to the O5-plane, while the dashes correspond to the directions in which the O5-planes are localised, at the values $\{0,1/2\}$.}
    \label{tab:O5sGammaHat1}
\end{table}\\
The key idea is that by appropriately choosing the vectors $\vec{a}$, $\vec{b}$, and $\vec{c}$, along with a suitable orientifold involution $\sigma$, one has considerable flexibility in generating intersecting O5-planes along various directions. 
While the discussion so far has focused on orbifolds of the ordinary seven-torus $\mathbb{T}^7$, these constructions can often be extended to twisted tori $\tilde{\mathbb{T}}^7$. In such cases, however, the lattice by which one quotients does not need to be the square lattice, which is indeed the situation for nilmanifolds.
When twisted tori are quotiented by any of the orbifold groups $\hat{\Gamma}$, the untwisted Betti numbers are reduced. Indeed, there are no one- or two-forms that are preserved under the orbifold, i.e. $b^1(\mathcal{M}) = b^2(\mathcal{M}) =0$, and there are only seven three-forms and seven four-forms that are preserved, i.e. $b^3(\mathcal{M})=7$.
A basis for them is given by 
\begin{align}
\label{eq:basis_three_four_forms}
\begin{split}
    &\Phi^i = \left\{ e^{127},\, e^{347},\, e^{567},\,e^{136},\, e^{235},\,e^{145},\,- e^{246} \right\}\,,\\
    &\Psi^i = \left \{ e^{3456},\, e^{1256},\, e^{1234},\, e^{2457},\, e^{1467},\, e^{2367},\, -e^{1357}\right\}\,.
\end{split}
\end{align}
where the short-hand notation $e^{a...b}=e^a \wedge ... \wedge e^b$ is used. 
Consequentially, the seven-dimensional manifold admits therefore a $G_2$-structure with a left-invariant three-form $\Phi$ and its associated four-form $\star \Phi$, which can be parametrised as follows:
\begin{align}
\label{eq:Phi_and_starPhi}
\begin{split}
    \Phi &= \; s_{127} \:e^{127}+s_{347}\: e^{347}+s_{567}\: e^{567} + s_{136}\: e^{136}+s_{235}\: e^{235} + s_{145}\: e^{145}- s_{246}\: e^{246}\,,\\
    \star\Phi &= \; s_{3456}\: e^{3456} + s_{1256}\: e^{1256}+s_{1234}\: e^{1234} + s_{2457}\: e^{2457} + s_{1467}\: e^{1467}\\
    & \quad + s_{2357}\: e^{2367} - s_{1357}\: e^{1357}\,,
\end{split}
\end{align}
where $s_{ijk} = L_i L_j L_k$ and $s_{ijkl} = L_i L_j L_k L_l$.
Again, the forms $\Phi$ and $\star \Phi$ are invariant under all orbifold actions $\hat{\Gamma}$, and they are even under the orientifold involution $\sigma_{\hat{\gamma}}$ as well.
The $G_2$ left-invariant three- and four-forms do generally not need to be closed. Their non-closure can be described by the torsion classes, classified in different irreducible representations of the $G_2$ group:
\begin{equation}
\label{eq:G2_Torsion}
\dd \Phi = W_1 \star \Phi + W_7 \wedge \Phi + W_{27}\,, \qquad\dd \star \Phi = \frac{4}{3} W_7 \wedge \star \Phi + W_{14} \wedge \Phi.
\end{equation}
The Ricci curvature can also be expressed in terms of the torsion classes:
\begin{equation}
    R_7  = \frac{21}{8}|W_1|^2+\frac{10}{3} |W_{7}|^2+ 4 \star \dd \star W_{7}- \frac{1}{2} |W_{14}|^2- \frac{1}{2} |W_{27}|^2\,.
\end{equation}
Finally, note that all seven four-forms \eqref{eq:basis_three_four_forms} are closed under the exterior derivative by the Maurer-Cartan equations \eqref{eq:MaurerCartan} and the allowed structure constants \eqref{eq:allowed_structure_constants}. Consequentially, this means that the $\mathbb{Z}_2^3$ orbifolds lead to co-closed $G_2$-structures:
\begin{equation}
\label{eq:def_co-closed_G2}
    \dd \Phi = W_1 \star \Phi + W_{27}\,, \qquad \dd \star \Phi = 0\,.
\end{equation}
Co-closed $G_2$-structures play an important role in minimally supersymmetric type IIB compactifications to AdS$_3$, as is discussed below. 

\section{Scale-separated solutions of type IIB with co-closed \texorpdfstring{$G_2$}{G2}-structures}
\label{sec:G2-solutions}
This section constructs AdS$_3$ solutions of type IIB string theory, for which co-closed $G_2$-structures are essential. Solutions of this type have been studied before in  refs.~\cite{Passias:2020ubv,Emelin:2021gzx} and \cite{VanHemelryck:2022ynr}, and the techniques used there apply directly here. Compactifications on these spaces preserve 1/8 of the original supersymmetry, and the inclusion of orientifold planes breaks half of the remaining supersymmetries, leading to minimal supersymmetry in three dimensions. The ten-dimensional metric can be written as a direct sum as follows:
\begin{equation}
    \dd s_{10}^2 = \e^{2A}\dd s_3^2 + \dd s_7^2\,,
\end{equation}
where $A$ is the warp factor, which is a function on the seven-dimensional space, governed by the metric $\dd s_7^2$, which measures distances in string units. To find supersymmetric solutions, one can solve the Killing spinor equations of type IIB string theory and impose the Bianchi identities.
The Killing spinor equations were rewritten in ref.~\cite{Dibitetto:2018ftj} in terms of bispinors for general backgrounds and applied in \cite{Passias:2020ubv,VanHemelryck:2022ynr} for $G_2$-structure backgrounds in type IIB. They require that the $G_2$-structure must be co-closed, a combination of the warping and the dilaton must be constant and that the NSNS three-form must vanish:
\begin{equation}
\label{eq:starPhiTorsion}
    \dd \star \Phi  =0\,, \qquad  H = 0\,, \qquad \e^{2A-\phi} = g_s^{-1}\,.
\end{equation}
Furthermore, the geometry is fixed by the RR fluxes, however, it is more convenient to take the other perspective and express the fluxes in terms of the geometry
\begin{align}
    &F_1 =  0\,,\\
\label{eq:F3BPS}
    &F_3 = - g_s^{-1}\e^{-3A}\star\dd \left(\e^{A}\Phi\right) + 2\mu g_s^{-1} \e^{-3A}\Phi\,,\\
    &F_5 = 0\,,\\
\label{eq:F7BPS}
    &F_7 = - 2\mu g_s^{-1} \; \vol_7\,.
\end{align}
The constant $\mu$ is related to the inverse AdS radius, i.e. $\mu^2 = 1/L_\mathrm{AdS}^2$. 
Finally, the torsion class $W_1$ and AdS scale are also related by the Killing spinor equations:
\begin{equation}
\label{eq:W1isMu}
    W_1 = \frac{12}{7} \mu\: \e^{-A}.
\end{equation}
The RR three-form flux must also satisfy its Bianchi identity, which can be sourced by D5-branes or O5-planes. Here, it is enough to consider O5-planes only, and the relevant Bianchi identity in string units becomes:
\begin{equation}
\label{eq:BI_F3}
    \dd F_3 = -\sum_{i,j} \delta(\Sigma_{\text{O5}_{\perp,i,j}})\,.
\end{equation}
The sum over $i$ runs over distinct O5-plane sets in different directions, while the sum over $j$ accounts for multiple O5-plane images parallel to those directions.
In the rest of this work, the smeared approximation for the O-planes is imposed, where the Dirac-delta forms are replaced by the unit volume forms of the cycles that the O5's are orthogonal to, i.e. $\sum_{j}\delta(\Sigma_{\text{O5}_{\perp,i,j}}) \to j_{\text{O5}_i} = N_\mathrm{O5} \vol(\Sigma_{\text{O5}_{\perp,i}})$ where unit volume is understood here and $N_\mathrm{O5}$ is the number of O5-planes. 
In that case, the warping and dilaton profile are constant, and the solution for $F_3$ reduces to 
\begin{equation}
\label{eq:F3smeared}
    g_s F_3 = \frac{W_1}{6} \Phi - \star W_{27}\,.
\end{equation}
One concludes that, for any co-closed $G_2$-structure manifold, there exists a supersymmetric 3d compactification, provided that the manifold has the proper orientifold content such that the $F_3$-Bianchi identity is satisfied (in the smeared approximation). 
As stated above, orbifolds of various seven-dimensional nilmanifolds and solvmanifolds are examples of co-closed $G_2$-structures and can therefore be used for these type IIB compactifications, which are discussed in the subsections below. 

\subsubsection*{Masses and conformal dimensions}
Before examining specific solutions, the masses of the scalar fields and the conformal dimensions of their putative holographic dual operators are discussed here.
In refs.~\cite{Conlon:2021cjk, Apers:2022tfm,Quirant:2022fpn}, it was noticed that in the parametrically scale-separated AdS$_4$ vacua of massive type IIA \cite{DeWolfe:2005uu, Camara:2005dc}, the masses of the scalars are such that their corresponding operators in the putative holographic field theory have integer conformal dimensions. Although originally analysed for the $\mathbb{Z}_3\times \mathbb{Z}_3$ orbifold of \cite{DeWolfe:2005uu, Camara:2005dc}, it was shown in ref.~\cite{Apers:2022zjx}, that the same phenomenon occurs for other toroidal orbifolds, including $\mathbb{Z}_2\times \mathbb{Z}_2$. Therefore, the same feature arises in the scale-separated vacua of ref.~\cite{Cribiori:2021djm}, as they are morally T-dual to the vacua of refs.~\cite{DeWolfe:2005uu, Camara:2005dc} on the $\mathbb{Z}_2\times \mathbb{Z}_2$ orbifold, as discussed in ref.~\cite{Apers:2022zjx}. In the three-dimensional vacua of ref.~\cite{Farakos:2020phe}, the situation is quite different, as it was shown in ref.~\cite{Apers:2022zjx} that the conformal dimensions are not integers there. Hence, it is worthwhile to investigate whether this feature also appears for the vacua constructed here.

As was explained well in ref.~\cite{Emelin:2021gzx}, there are no axions in the type IIB setup here, either because homology does not allow for it (e.g. there is no $B$ or $C_2$ axion as there are no two-forms in the untwisted sector) or because the orientifold projects them out (e.g. $C_4$ is odd under an O5-involution but there are no odd four-cycles in the manifold). So the only scalar fields that enter the potential are the seven three-cycle volumes (or equivalently, the seven radii) and the dilaton.
Since the radii $L_k$ defined above parametrise volumes in string units, it is convenient to trade the 10d dilaton $\phi$ for the 3d dilaton $\delta$. Additionally, it is useful here to trade the three-cycle volume scalars for the radii, such that the field space metric is diagonal:
\begin{equation}
    g_{ij}\partial \sigma^i \partial \sigma^j =  4\frac{(\partial \delta)^2}{\delta^2}+\sum_{k=1}^{7}\frac{(\partial L_k)^2}{L_k^2} \,,
\end{equation}
where the relation between the dilatons is that
\begin{equation}
    \delta^2 = \e^{-2\phi}L_1 L_2 L_3 L_4 L_5 L_6 L_7\,.
\end{equation}
The scalar potential was already obtained from dimensional reduction in type IIB in ref.~\cite{Emelin:2021gzx} and is given by:\footnote{Note that there is a difference of a factor 4 for the first term of the potential when comparing with ref.~\cite{Emelin:2021gzx}, which is a result of normalising the kinetic terms differently.}
\begin{equation}
\label{eq:V_pot}
    V = 4\left(g^{ij} (\partial_i P) (\partial_j P) - P^2\right)\,,
\end{equation}
where $g_{ij}$ is the field space metric and $P$ is the real superpotential first defined in ref.~\cite{Emelin:2021gzx} and derived in terms of bispinors in ref.~\cite{VanHemelryck:2022ynr}.
It is given by
\begin{equation}
\label{eq:P_pot}
    P = \e^{\mathcal{K}/2} \mathcal{W}\,,
\end{equation}
where the functions $\mathcal{K}$ and $\mathcal{W}$ are determined by the fluxes and geometry:
\begin{equation}
\label{eq:WK_pot}
    \mathcal{W} = -\frac{1}{4} \int \e^{-\phi}\left( \star \Phi \wedge F_3 - F_7\right) +\frac{1}{8} \int \e^{-2\phi}\left( \Phi \wedge \dd \Phi\right)\,, \quad \mathcal{K} = -4 \log \left[\int \e^{-2\phi}\vol_7 \right]\,.
\end{equation}
The supersymmetry conditions \eqref{eq:starPhiTorsion}-\eqref{eq:F7BPS} are equivalent to the condition that $\partial_i P=0$ for all eight scalar fields.
Again, it is more straightforward to calculate the mass matrix by trading $\phi$ for $\delta$. The masses are usually expressed in AdS units, therefore, it is useful to observe that $V = \Lambda = -2 L_\mathrm{AdS}^{-2}$ and hence 
\begin{equation}
\label{eq:general_masses}
    m^2 L_\mathrm{AdS}^2 = \mathrm{Eig}(M)\,, \qquad M^{ij} = \left(-\frac{2}{V}\right) \sqrt{g}^{ik}(\partial_{kl}V)\sqrt{g}^{lj}\,.
\end{equation}
The conformal dimensions of the field theory operators dual to the scalars are given by the simple relation
\begin{equation}
    \Delta = 1 + \sqrt{1+ m^2 L_\mathrm{AdS}^2}\,, \qquad m^2 L_\mathrm{AdS}^2 = \Delta(\Delta-2)\,.
\end{equation}
In the next sections, specific solutions on solv- and nilmanifolds are discussed.

\subsection{Solutions on solvmanifolds}
Now that all necessary material on type IIB AdS$_3$ compactifications on co-closed $G_2$-structures is introduced, it is time to focus on solutions with a particular example of a seven-dimensional solvmanifold. Solvmanifolds are group manifolds of a solvable group, which are subgroups of the group of upper-triangular matrices. Another defining characteristic is that the derived series of the Lie algebra $\mathfrak{g}$, which is recursively defined as the commutant $\mathfrak{g}_k \equiv [\mathfrak{g}_{k-1},\mathfrak{g}_{k-1}]$ with $\mathfrak{g}_1 \equiv \mathfrak{g}$, terminates for some $k$ at $\mathfrak{g}_k = \{0\}$. The particular seven-dimensional solvmanifold of interest here is characterised by the following Maurer-Cartan equations:
\begin{align}
\label{eq:MaurerCartanSolv}
\begin{split}
    &\dd e^1 = \omega_a\; e^2 \wedge e^7, \qquad \dd e^2 = - \; \omega_a e^1 \wedge e^7\,,\\
    &\dd e^3 = \omega_b\; e^4 \wedge e^7, \qquad \dd e^4 = - \; \omega_b e^3 \wedge e^7\,,\\
    &\dd e^5 = \omega_c\; e^6 \wedge e^7, \qquad \dd e^6 = - \; \omega_c e^5 \wedge e^7\,.
\end{split}
\end{align}
These can be integrated into the following expressions for the one-forms, in terms of the  coordinates $x^1,...,x^7$:
\begin{align}
\label{eq:one-forms_in_coords}
\begin{split}
    &e^1 = \cos\left(\omega_a x^7\right) \dd x^1 - \sin\left(\omega_a x^7\right)\dd x^2\,,\\
    &e^2 = \sin\left(\omega_a x^7\right) \dd x^1 + \cos\left(\omega_a x^7\right) \dd x^2\,,\\
    & e^7 = \dd x^7\,.
    \end{split}
\end{align}
and for the other directions idem upon the simultaneous cyclic permutation of (1,3,5), (2,4,6) and $(\omega_a,\omega_b, \omega_c)$. This manifold has been considered before, for example, in refs.~\cite{DallAgata:2005zlf,Andriot:2015sia, Manero2020315}.
In essence, this seven-dimensional solvmanifold can be seen as three different copies of the 3d solvmanifold $E_2$, also called ISO(2), with one common direction (the seventh), see e.g. refs.~\cite{DallAgata:2005zlf,Grana:2006kf,Grana:2013ila,Andriot:2015sia}.
The three-dimensional space can be divided by different lattices to make it compact, each lattice leading to a different cohomology, see refs.~\cite{Grana:2013ila,Andriot:2015sia}. However, for this paper, it is enough to take the standard torus lattice $x^i \simeq x^i + 1$, as the Betti numbers are reduced through an orbifold.
Consequently, this means that the structure constants $\omega_{a,b,c}$ have to be quantised by $2\pi\mathbb{Z}$ since the coframe has to be preserved by the lattice. The orbifolds $\hat\Gamma^4$ and $\hat \Gamma^1$ can be imposed on this solvmanifold as well, and by orientifolding with $\sigma_\gamma$, one obtains the same orientifold planes as in Tables \ref{tab:O5sGammaHat4} and \ref{tab:O5sGammaHat1}. For these orbifolds of the ordinary torus, it was argued that the singularities can be cured by a blow-up procedure. Although it is not straightforward to extrapolate this to the twisted torus $\tilde{\mathbb{T}}^7$ here, one could argue that, in a neighbourhood of the fixed points of $\hat{\gamma}$, the twisted torus looks locally like the ordinary one. For $\hat{\alpha}$ on the other hand, the fixed loci are the three-dimensional solvmanifolds ISO(2) instead of $\mathbb{T}^3$'s. However, if one would choose to set $\omega_b=0$, then they are $\mathbb{T}^3$'s, and the arguments above might apply again, although one has to treat this reasoning with caution. 
For further purposes, it is also interesting to look at the curvature of this space, which can be computed from the metric and structure constants: 
\begin{align}
    &R_7 = - \frac{1}{2L_7^2}\left[\omega_a^2\left(\frac{L_1}{L_2}-  \frac{L_2}{L_1}\right)^2+\omega_b^2\left(\frac{L_3}{L_4}-  \frac{L_4}{L_3}\right)^2+\omega_c^2\left(\frac{L_5}{L_6}-  \frac{L_6}{L_5}\right)^2\right]\,,\\
    &R_{77} = \frac{1}{L_7^2} R_7\,, \qquad \frac{1}{L_1^2} R_{11} = -\frac{1}{L_2^2} R_{22} = \frac{\omega_a^2}{2 L_7^2} \left(\frac{L_1^2}{L_2^2}-\frac{L_2^2}{L_1^2}\right)\,, \quad \mathrm{etc.}
\end{align}
where the Ricci tensor components are expressed in flat indices corresponding to the metric \eqref{eq:metric}, and the other components involving directions 3, 4, 5 and 6 are analogous under the simultaneous replacement of $(1,2) \leftrightarrow (3,4) \leftrightarrow (5,6)$  and $a \leftrightarrow b \leftrightarrow c$. Additionally, the torsion class $W_1$ takes the following form:
\begin{equation}
\label{eq:solv_W1}
    W_1 =\frac{2}{7 L_7}\left[\omega_a\left(\frac{L_1}{L_2}+  \frac{L_2}{L_1}\right)+\omega_b\left(\frac{L_3}{L_4}+  \frac{L_4}{L_3}\right)+\omega_c\left(\frac{L_5}{L_6}+  \frac{L_6}{L_5}\right)\right]\,.
\end{equation}

\subsubsection*{Solutions with four intersecting O5-planes}
First, consider the space the solvmanifold orbifolded by $\hat{\Gamma}_4$ of section~\ref{sec:orbifolds}, which, upon orientifolding, admits four sets of orientifold planes as in Table~\ref{tab:O5sGammaHat4}. It turns out to be useful to look at the situation where $L_1 = L_2$,  $L_3 = L_4$ and $L_5 = L_6$, which simplifies the setup but such that not all three-cycle volumes are the same. For this choice, the exterior derivative on $\Phi$ gives the simple expression:
\begin{equation}
\label{eq:dPhi_solv_4}
    \dd \Phi = L_1 L_3 L_5 \omega_\Sigma \left(e^{2457} + e^{1467} + e^{2367} - e^{1357} \right), \quad \omega_\Sigma = \omega_a + \omega_b + \omega_c\,, \quad W_1= \frac{4\omega_\Sigma}{7 L_7}\,.
\end{equation}
Note that the torsion class $W_{27}$ can also be obtained by comparing eq.\eqref{eq:dPhi_solv_4} with eqns.~\eqref{eq:Phi_and_starPhi} and \eqref{eq:def_co-closed_G2}. When $\omega_\Sigma$ vanishes, the $G_2$-structure is both closed and co-closed. This was already noticed in refs.~\cite{Andriot:2015sia,Manero2020315}. In that case, $\mu$ also vanishes due to eq.~\eqref{eq:W1isMu} and the external spacetime is Minkowski instead of Anti-de Sitter. So for now, one can take $\omega_\Sigma \neq 0$. Even in this case, the Ricci curvature $R_7$ vanishes, and even more so, the manifold is Ricci-flat, i.e. $R_{mn}=0$.
Furthermore, the supersymmetric solutions above, eqs.~\eqref{eq:F3BPS}-\eqref{eq:F7BPS} determine $F_3$ and $F_7$ completely in terms of the geometry:
\begin{gather}
\label{eq:F3sol_GammaHat}
    F_3 =  \left(f_{3,1}\;e^{127} + f_{3,3}\;e^{347} + f_{3,5}\;e^{567} \right) - \tilde{f}_3\left(e^{136}+e^{235}+e^{145}-e^{246} \right), \quad  F_7 = -f_7\, e^{1234567},\\
    f_{3,i}=\frac{2 L_i^2 \omega_\Sigma}{3 g_s }\,, \qquad \tilde{f}_3 = \frac{L_1 L_3 L_5 \omega_\Sigma}{3 g_s L_7}\,, \qquad f_7 = \frac{2\omega_\Sigma (L_1 L_3 L_5)^2}{3 g_s}\,,
\end{gather}
where the index $i$ runs over $\{1,3,5\}$. The radii and string coupling can also be expressed in terms of the flux parameters by inverting the last identities:
\begin{equation}
    L_i = \left(\frac{f_7 \;f_{3,i}^2}{f_{3,1}\;f_{3,3}\;f_{3,5}}\right)^{\frac{1}{4}} , \quad L_7 = \frac{\left(f_7\; f_{3,1}\;f_{3,3}\;f_{3,5}\right)^{\frac{1}{4}}}{2\tilde{f}_3}, \quad g_s = \frac{2 \omega_\Sigma}{3}\left(\frac
    {f_7}{f_{3,1}\;f_{3,3}\;f_{3,5}}\right)^{\frac{1}{2}} \,.
\end{equation}
Furthermore, $F_3$ is not closed, but due to the Maurer-Cartan equations \eqref{eq:MaurerCartanSolv}, only the components including $\tilde{f}_3$ are responsible for that:
\begin{equation}
    \dd F_3 = -\tilde{f}_3 \omega_\Sigma \left(e^{2457} + e^{1467} + e^{2367} - e^{1357} \right)\,,
\end{equation}
and since $ \tilde{f}_3 \omega_\Sigma > 0$, the Bianchi identity of $F_3$ \eqref{eq:BI_F3} should be solved by four sets of O5-planes. And indeed, as discussed in the previous sections, by orientifolding $\tilde{\mathbb{T}}^7/\hat{\Gamma}_4$, one gets, upon smearing, precisely the O5-planes required as in Table~\ref{tab:O5sGammaHat4}.
Note also that the components of $F_3$ including the $f_{3,i}$'s are closed and do not participate in the Bianchi identity. This implies that all the $f_{3,i}$'s are unbounded flux quanta that can be chosen to be arbitrarily large.
Finally, the masses and conformal dimensions, by eq.~\eqref{eq:general_masses}, are determined to be:
\begin{align}
    m^2 L_\mathrm{AdS}^2 &= \left\{3,8,8,8,8,\left(6\frac{\omega_a}{\omega_\Sigma}\right)^2-1,\left(6\frac{\omega_b}{\omega_\Sigma}\right)^2-1,\left(6\frac{\omega_c}{\omega_\Sigma}\right)^2-1\right\}\,,\\
    \Delta &= \left\{3, 4, 4, 4, 4, 1 + 6\frac{|\omega_a|}{|\omega_\Sigma|}, 1 + 6\frac{|\omega_b|}{|\omega_\Sigma|}, 1 + 6\frac{|\omega_c|}{|\omega_\Sigma|} \right\}\,.
\end{align}
So the conformal dimensions are all integer when $|\omega_{a,b,c}/ \omega_\Sigma|$ are elements of $\mathbb{Z}/6$. This happens for the most symmetric case, $\omega_a = \omega_b = \omega_c$, but also for $-\omega_a = \omega_b = \omega_c$, for example. Note that for these cases, all masses are non-negative as well, and hence all geometric moduli are stabilised. The tadpole constraint only limits the size of $\omega_\Sigma = \omega_a + \omega_b + \omega_c$, still allowing for various spectra.

\subsubsection*{Solutions with one set of O5-planes}
Another interesting option is to use the orbifold $\hat{\Gamma}_1$ from \eqref{eq:orbifold_Gamma1}, where the radii are also related in pairs by $L_{2} = \sqrt{2}L_{1}$, $L_{4} = \sqrt{2}L_{3}$ and $L_{6} = \sqrt{2}L_{5}$. Additionally, it is interesting to choose all structure constants to be the same in size, but where one of them has the opposite sign, which can be taken to be $\omega_a$ without loss of generality, i.e. $-\omega_a = \omega_b = \omega_c$. For this choice, the non-closure of $\Phi$ is captured by:
\begin{equation}
    \dd \Phi = - \sqrt{2}L_1 L_3 L_5 \omega_a \left(2e^{2367}+2 e^{2457}- e^{1357} \right), \qquad W_1= -\frac{3 \sqrt{2} \omega_a}{7 L_7}\,,
\end{equation}
and the scalar curvature does not vanish. Using this space, the RR three-form flux is of a similar form as above:
\begin{gather}
    F_3 =  -\left(f_{3,1}\;e^{127} + f_{3,3}\;e^{347} + f_{3,5}\;e^{567} \right) + \tilde{f}_3\left(e^{136}-e^{235}+e^{145}-2 e^{246} \right), \quad  F_7 = f_7\, e^{1234567},\\
    f_{3,i}=\frac{ L_i^2 \omega_a}{g_s }\,, \qquad \tilde{f}_3 = \frac{L_1 L_3 L_5 \omega_a}{g_s L_7}\,, \qquad f_7 = \frac{2\omega_a (L_1 L_3 L_5)^2}{g_s}\,,
\end{gather}
where the index $i$ runs again through $\{1,3,5\}$. Expressing the scalars in terms of the fluxes gives:
\begin{equation}
    L_i = \left(\frac{f_7\;f_{3,i}^2}{2 f_{3,1}\;f_{3,3}\;f_{3,5}}\right)^{\frac{1}{4}}, \quad L_7 = \frac{\left(f_7\;f_{3,1}\;f_{3,3}\;f_{3,5}\right)^{\frac{1}{4}}}{ 2^{\frac{1}{4}}\tilde{f}_3}, \quad g_s = \omega_a \left(\frac{f_7}{2f_{3,1}\;f_{3,3}\;f_{3,5}}\right)^{\frac{1}{2}} \,.
\end{equation}
Note that there are some similarities but also key differences with the solution \eqref{eq:F3sol_GammaHat} on the previous orbifold. For example, the signs are different, especially the one for the 235-component, and the 246-component has an additional factor of 2. Nevertheless, the dependence on the flux parameters for the radii and string coupling is the same as what was considered for $\tilde{\mathbb{T}}^7/\hat{\Gamma}_4$.
The RR three-form is not closed here either, and again, only the components including $\tilde{f}_3$ are responsible for this, but this time they only generate one component:
\begin{equation}
    \dd F_3 = -3 \tilde{f}_3\, \omega_a \, \left(-e^{1357}\right)\,.
\end{equation}
Hence, only one set of (smeared) O5-planes is required to satisfy the Bianchi identity. As indicated above, orientifolding $\tilde{\mathbb{T}}^7/\hat{\Gamma}_1$ generates only one set of O5-planes along the proper directions, as in Table~\ref{tab:O5sGammaHat1}. Again, the $f_{3,i}$'s are not restricted by the tadpole and can be chosen to be arbitrarily large.
Note that solutions on this space should be seen as the supersymmetric cousins of some of the solutions found in ref.~\cite{Arboleya:2024vnp,Arboleya:2025ocb}, which also include just one set of O5-planes and are non-supersymmetric. However, that paper uses another orbifold than the specific $\tilde{\mathbb{T}}^7/\hat{\Gamma}_1$ and therefore might need additional D5-branes to cancel O5-charge along other directions.

For the models on $\tilde{\mathbb{T}}^7/\hat{\Gamma}_1$, the masses and corresponding conformal dimensions are:
\begin{equation}
    m^2 L_\mathrm{AdS}^2 = \{48, 48, 48, 8, 8, 8, 8, 0\}\,, \qquad \Delta= \{8, 8, 8, 4, 4, 4, 4, 2\}\,.
\end{equation}
Hence, one of the scalars is massless but does not seem to correspond to a flat direction of the potential as the scalars are fixed, similar to refs.~\cite{Becker:2024ayh,Becker:2024ijy,Rajaguru:2024emw}.
Notice that this spectrum is contained within the spectra of the non-supersymmetric scale-separated vacua of ref.~\cite{Arboleya:2024vnp,Arboleya:2025ocb}. That is quite surprising, as the authors there were looking at similar vacua with one O5-plane but with an SO(3)-invariant Ansatz that would require all structure constants $\omega_{a,b,c}$ to be the same (similar for all $f_{3,i}$'s). However, taking that choice here, but also artificially changing the fluxes $F_{246} \to  - F_{246}$, $F_{235}= - F_{235}$, would amount to breaking supersymmetry, but corresponds to the solution found in ref.~\cite{Arboleya:2024vnp,Arboleya:2025ocb}. Applying these changes does result in the same expression for the scalar potential after carefully choosing $F_{136}= -F_{235}=F_{145}=-2F_{246}$ and $-\omega_a = \omega_b = \omega_c$ and hence can be understood as a ``skew-whiffing'' procedure \cite{Duff:1984sv,Duff:1986hr}, although breaking with the SO(3)-invariant Ansatz.

\subsubsection*{Scale separation?}
Now it is time to investigate whether there is a separation of scales between the AdS scale and the volume of the three-cycles of the manifold. For this purpose, it is essential to make use of the fluxes that are unrestricted by tadpoles, which are parametrised here by the $f_{3,i}$'s and $f_7$. The flux parameter $\tilde{f}_3$ and structure constants $\omega_{a,b,c}$ are constrained by the $F_3$ Bianchi identity, as they have to be balanced against the O5-plane charge density. The latter is fixed, hence it cannot be tuned to infinity.
Indeed, for both models, the Bianchi identity imposes that the product $\tilde{f}_3\omega_\Sigma$ for $\tilde{\mathbb{T}}^7/\hat{\Gamma}_4$ or $\tilde{f}_3 \omega_a$ for $\tilde{\mathbb{T}}^7/\hat{\Gamma}_1$ should be held fixed. However, all flux parameters are quantised, also $\tilde{f}_3$ and $\omega_\Sigma$ or $\omega_a$, which means that they should be held fixed separately: neither one of them can be made arbitrarily small at the expense of the other one becoming large.
Another important remark is that the AdS scale is of similar size to the radius $L_7$, as  
\begin{equation}
    L_\mathrm{AdS}^{-1} = |\mu| = \frac{7}{12}|W_1| \sim \omega_\Sigma L_7^{-1} \text{ or }\omega_a L_7^{-1}\,.
\end{equation}
However, to achieve scale separation with the cycle volumes, this is not a problem, as there are no cycles whose volumes are determined by $L_7$ alone, as one-forms are projected out by the orbifold.
The volumes of the generalised three-cycles are then set by $L_i^2 L_7$ and $L_1 L_3 L_5$ for $i \in \{1,3,5\}$, which satisfy
\begin{equation}
    \frac{L_i^2 L_7}{L_\mathrm{AdS}^3} \sim \frac{f_{3,i}}{f_{3,1}\;f_{3,3}\;f_{3,5}}\,, \qquad \frac{L_1 L_3 L_5}{L_\mathrm{AdS}^3} \sim \frac{1}{f_{3,1}\;f_{3,3}\;f_{3,5}}\,,
\end{equation}
and indeed, by sending the $f_{3,i}$'s to infinity, one obtains parametric scale separation.
To ensure that all cycles are large in string units and that the string coupling is small, one should take the $f_{3,i}$'s and $f_7$ large, such that
\begin{equation}
    \frac{f_{3,1}\;f_{3,3}\;f_{3,5}}{f_{3,\alpha}^2} \ll f_7 \ll f_{3,1}\;f_{3,3}\;f_{3,5}\,.
\end{equation}
where $f_{3,\alpha}$ is the smallest flux amongst the $f_{3,i}$'s.
Hence, scale separation \textit{with the cycle volumes} can be achieved in the controlled regime, i.e.~at small string coupling and large volume. 
However, it has to be clarified that \textit{this does not guarantee a hierarchy between the AdS- and the KK radius as given by the lowest non-trivial eigenvalue of the scalar Laplacian} necessarily. Indeed, although the radius $L_7$ does not parametrise a volume on its own, it seems to determine a tower of eigenvalues of the Laplacian.  Indeed, there is a set of eigenfunctions of the Laplacian operator that do not depend on the coordinates $x^{1,2,3,4,5,6}$, and are just the eigenfunctions on a circle with radius $L_7$. This is discussed in more detail in appendix \ref{app:solv_spectrum}. It is not expected that the $\mathbb{Z}_2^3$ orbifold projects out an entire tower of modes modes, but the decisive answer is unclear, as orbifolds can change the eigenvalue spectrum. This (non-)isospectrality problem is an actively studied topic in mathematics, see refs.~\cite{MR739790,MR4767506,MR4501837} for a very select overview. Nevertheless, the non-isospectrality of orbifolds lies typically in a different multiplicity of eigenvalues, but not necessarily so in some distinct eigenvalues being projected out entirely. Exploring whether the latter could happen for the solvmanifolds under consideration here, is left for future research.
 
With all these considerations, it looks like one can only obtain scale separation when $W_1$ really decouples from all the radii. Looking at eq.~\eqref{eq:solv_W1}, this requires a fine-tuned situation in which not all $\omega$'s have the same sign and the radii only differ slightly as to render $L_7 W_1$ small. However, this should be compatible with the supersymmetry equations, flux quantisation and the tadpole, and hence it is unclear whether this can happen. 
 
If one would nevertheless be interested in holographic aspects of this solvmanifold compactification, then it is interesting to see what the central charge of the putative holographic dual theory looks like. For that, one needs to know how the 3d Planck mass scales with the fluxes when expressed in string units, here given by
\begin{equation}
    \Mpl = g_s^{-2} \mathcal{V}_7 \sim (f_7\;f_{3,1}\;f_{3,3}\;f_{3,5})^{\frac{3}{4}}\,.
\end{equation}
Here $\mathcal{V}_7= L_1 L_2 L_3 L_4 L_5 L_6 L_7$ is the volume of the compact space.
This would lead in a putative holographic dual theory to a central charge that satisfies:
\begin{equation}
    c = \frac{2}{3}\Mpl L_\mathrm{AdS} \sim f_7 \; f_{3,1}\;f_{3,3}\;f_{3,5} \implies f_7^2 \ll c \ll f_7^2 f_{3,\alpha}^2\,.
\end{equation}
This would result in a relatively large number of degrees of freedom in the putative holographic dual theory. Additionally, there would be a parametrically large gap in the spectrum. Such conformal field theories with this feature have not been found yet.

\subsection{Solutions on nilmanifolds}
The seven-dimensional solvmanifolds from the last subsection provide solutions at weak coupling, large volumes and scale separation with those volumes, but not necessarily with the eigenvalues of the scalar Laplacian. The technical reason for this is that the torsion class $W_1$, which determines the inverse AdS radius through the supersymmetry equations \eqref{eq:W1isMu}, does not decouple from all radii in the large flux regime, and those radii typically determine the eigenvalues of the Laplacian and hence the KK mass scales. 
For nilmanifolds, these issues do not necessarily arise. Nilmanifolds are group manifolds where the characterising group is a subgroup of the group of upper-triangular matrices excluding the diagonal. Another defining characteristic of nilmanifolds is that the lower central series of the Lie algebra $\mathfrak{g}$, now defined by $\mathfrak{g}^k \equiv [\mathfrak{g},\mathfrak{g}^{k-1}]$ with $\mathfrak{g}^0=\mathfrak{g}$, terminates for some $k$ at $\mathfrak{g}^k = \{0\}$, which are then called `$k$-step nilmanifolds'.
Seven-dimensional nilmanifolds have been classified in ref.~\cite{MR2698220}, and those that admit a co-closed $G_2$-structure have been classified in \cite{MR2811660,MR3739330,MR4626831,MR4789076}. However, it turns out that only 7 out of 24 decomposable nilmanifolds (i.e. the space can be written as a product of a lower-dimensional nilmanifold and an ordinary torus) are consistent with the $\mathbb{Z}_2^3$ orbifolds under consideration. Of the indecomposable ones, only 5 two-step nilmanifolds are consistent with the orbifold, whereas for higher-step nilmanifolds are excluded by the orbifold \cite{DallAgata:2005zlf}. The algebras of these nilmanifolds are summarised in Table \ref{tab:nilmanifolds}, which also includes the directions in which O5-planes charges are necessary to cancel the tadpole. 
\begin{table}[ht]
    \centering
    \begin{tabular}{|l|l|l|c|} \hline
         Name & Algebra & O5-planes & WCSS \\ \hline
        $\mathfrak{n}_1$ & $(0,0,0,0,0,0,0)$ & /& $\times$ \\ \hline
        $\mathfrak{n}_2$ & $(0,0,0,0,0,0,12)$ & $\perp 1256, 1234$ & \checkmark\\ \hline
        $\mathfrak{n}_3$ & $(0,0,0,0,0,0,12+34)$ & $\perp 3456, 1256, 1234$ & $\times$\\ \hline
        $\mathfrak{n}_4$ & $(0,0,0,0,0,13,12)$ & $\perp  1357, 1256,  1234$ & \checkmark \\ \hline
        $\mathfrak{n}_6$ & $(0,0,0,0,0,13-24,12+34)$ & $\perp 1357, 2457, 3456, 1256, 1234^*$ & \checkmark\\ \hline
        $\mathfrak{n}_7$ & $(0,0,0,0,0,13,12+34)$ & $\perp 1357, 3456, 1256, 1234$ & $\times$ \\ \hline
        $\mathfrak{n}_{10}$ & $(0,0,0,0,23,13,12)$ & $\perp 1357, 2367, 1256, 1234$ & $\times$ \\ \hline \hline
        17 & $(0,0,0,0,0,0,12+34+56)$ & $\perp 3456, 1256, 1234$& $\times$\\ \hline
        37A & $(0,0,0,0,23,-24,12)$ & $\perp 2367, 2457, 1256, 1234$ & $\times$\\ \hline
        37B$_1$ & $(0,0,0,0,14,13-24,12+34)$ & $\perp 1357, 1467, 2457, 3456, 1256, 1234^*$ & \checkmark \\ \hline
        37C & $(0,0,0,0,23,-24,12+34)$ & $\perp 2367, 2457, 3456, 1256, 1234^*$  & \checkmark \\\hline
        37D$_1$ & $(0^4,14+23,13-24,12+34)$ & $\perp 1357, 1467, 2367, 2457, 3456, 1256, 1234^*$& \checkmark\\ \hline
    \end{tabular}
    \caption{All possible nilmanifolds that are consistent with the $\mathbb{Z}_2^3$ orbifold. In the second column, the entries encode the non-vanishing structure constants $f^{a}{}_{bc}$ of the algebra, with the upper index corresponding to place and the lower ones to the entries. e.g. for $\mathfrak{n}_3$, the non-vanishing structure constants are $f^{7}{}_{12}$ and $f^{7}{}_{34}$. Note that other parametrisations are possible, for example for $\mathfrak{n}_3$ having non-vanishing $f^{6}{}_{13}$ and $f^{6}{}_{24}$ instead, which lead to the same results. The third column illustrates the transverse directions of the O5-planes. An asterisk means that the O-plane in those directions can be present, but must be absent for well-controlled scale-separated solutions, which are indicated in the last column.}
    \label{tab:nilmanifolds}
\end{table}
The algebras are parametrised such that directions 5, 6 and 7 always belong to the centre of the algebra, meaning that only directions 5,6, and 7 are twisted and that their twists involve only directions 1, 2, 3 and 4 (nilmanifold 17 being the exception). 
For all nilmanifolds in the list (except nilmanifold 17), one can write the Maurer-Cartan equations as follows:
\begin{align}
\begin{split}
    &\dd e^{5} = f^{5}{}_{14}e^{14}+f^{5}{}_{23}e^{23}\,,\qquad \dd e^{6} = f^{6}{}_{13}e^{13}+f^{6}{}_{24}e^{24}\,,\\ &\dd e^{7} = f^{7}{}_{12}e^{12}+f^{7}{}_{34}e^{34}\,, \qquad \dd e^{1,2,3,4} = 0\,,
\end{split}
\end{align}
where one should put some of the structure constants to zero according to the algebra of choice in Table \ref{tab:nilmanifolds}. 
There are many different ways to integrate these expressions into coordinates. A specific choice is the following:
\begin{align}
\begin{split}
   &e^5 = \dd x^5 - f^{5}{}_{14} x^4 \dd x^1 - f^{5}{}_{23} x^3 \dd x^2\,, \qquad e^{6} = \dd x^6 - f^{6}{}_{13} x^3 \dd x^1 - f^{6}{}_{24}x^4 \dd x^2\,,\\
    &e^{7} = \dd x^7 - f^{7}{}_{12}x^2 \dd x^1 - f^{7}{}_{34} x^4 \dd x^3\,, \qquad e^{1,2,3,4} = \dd x^{1,2,3,4}\,.
\end{split}
\end{align}
The only thing that remains to make the nilmanifolds compact is to divide by a lattice. The following coordinate identifications do precisely that:
\begin{align}
\label{eq:nil_lattice}
    &x^{1,2,3,4} \simeq x^{1,2,3,4} + 2 m^{1,2,3,4}\,, \qquad x^5 \simeq x^5 + 2m^5 + 2 m^4 f^{5}{}_{14} x^1 + 2 m^3 f^{5}{}_{23} x^2\,,\\
    &x^{6} \simeq x^6 + 2 m^6 + 2 m^3 f^{6}{}_{13} x^1 +2 m^4 f^{6}{}_{24} x^2\,, \quad x^{7} \simeq x^7 + 2 m^7 + 2m^2 f^{7}{}_{12} x^1 +2 m^4 f^{7}{}_{34} x^3\,,
\end{align}
where $m^i \in \mathbb{Z}$ and all $f^{a}{}_{bc}\in \mathbb{Z}$. Notice that the lattice identifications leave the one-forms invariant. Additionally, the periodicity of the coordinates is now 2 instead of 1, as it was observed that under orientifolding, one would miss half of the O-plane images if the periodicity was kept to be 1, see Appendix D of ref.~\cite{Cribiori:2021djm}.
To see in which directions one needs O5-planes, one should compute the exterior derivative on $F_3$. 
Before doing so, it is useful to look at how the torsion class $W_1$ and the scalar curvature $R_7$ are expressed in terms of the structure constants:
\begin{equation}
\label{eq:general_W1_nil}
    W_1 = \frac{2}{7} \sum_{a, 1\leq b<c\leq 7}{\left(\sigma_{abc}f^a{}_{bc}\frac{L_a}{L_b L_c}\right)}\,, \qquad R_7 = - \sum_{a, 1\leq b<c\leq 7}{\frac{1}{2}\left(f^a{}_{bc}\frac{L_a}{L_b L_c}\right)^2}\,,
\end{equation}
where $\sigma_{abc} =+1$ for all $a,b,c$ except $\sigma_{624}=-1$, which arises from the fact that the 246-component of the three-form $\Phi$ comes with a minus sign. Important to note is that the curvature is always negative.
Due to the specific parametrisation of the algebras and their structure constants, the supersymmetric expression for $F_3$ \eqref{eq:F3smeared} can be written as follows:
\begin{equation}
	F_3 =  -\frac{7}{3} \frac{L_5 L_6 L_7}{g_s}  e^{567} + \sum_{a; b<c;f<g}\frac{L_f L_g L_a }{g_s}\left( \frac{7}{6}W_1\sigma_{a fg}-f^{a}{}_{bc} \frac{L_a }{L_b L_c}\; \epsilon^{bc}{}_{fg} \right)\; e^{fga}\,,
\end{equation}
where it is convenient to write out the 567 component separately. 
To see in which directions one needs O5-planes, one should compute the exterior derivative on $F_3$, which collapses to
\begin{equation}
\label{eq:dF3_nilmanifolds}
	 \dd F_3 = \underbrace{-\frac{7}{3}  g_s^{-1}(L_5 L_6 L_7)W_1}_{F_{567}} \frac{3}{2} f^{[5}{}_{bc}e^{67]}\wedge e^{bc}+ \underbrace{g_s^{-1}(L_1 L_2 L_3 L_4)\left(2 R_7 + \frac{49}{12} W_1^2\right)}_{4\: f^{a}{}_{[12} F_{34]a}} e^{1234}\,,
\end{equation}
where the first term involves multiple components but all coming from the 567-component of $F_3$, and the second term only contains the 1234-component. Due to eq.~\eqref{eq:general_W1_nil} and the Cauchy-Schwarz inequality and eq.~\eqref{eq:general_W1_nil}, the combination $(2 R_7 + 49 W_1^2/12)$ is either 1) negative definite or 2) zero. In case 1), it has to be cancelled with O5-plane charge, which cannot be tuned. Hence, the fluxes involved are restricted, which leads to either of the two following scenarios: 1a) it is not too constraining and there are still flux limits leading to large radii and weak string coupling, or 1b) it cannot.
In case 2), the supersymmetry conditions allow for 2a) only the combination to vanish or 2b) all individual fluxes involved to vanish. In case 2a), these fluxes are non-trivially unbounded, as they have to satisfy a constraint but can all be tuned to infinity.
Looking at the first term in \eqref{eq:dF3_nilmanifolds}, the algebra is such that each structure constant is responsible for distinct components in the first term of eq.~\eqref{eq:dF3_nilmanifolds}. Depending on the relative sign between the structure constants and the torsion class $W_1$, the tadpole has to be cancelled by D5-branes or O5-planes.
In the rest of this paper, only the presence of O5-planes is invoked. Looking at the first term of eq.~\eqref{eq:dF3_nilmanifolds}, and remembering that the total O5-plane charge is the same across different directions, this imposes that all non-vanishing structure constants to be the same in size, and more precisely,
\begin{equation}
\label{eq:O5_structure_constants}
	  f^{5}{}_{14}\,, f^{5}{}_{23}\,, f^{6}{}_{13}\,, -f^{6}{}_{24}\,, f^{7}{}_{12}\,, f^{7}{}_{34} \in \{0, \omega\}\,,
\end{equation}
where $\omega= f^{7}{}_{12}$ due to the parametrisation of the algebras.

Since the nilmanifolds are parametrised such that $f^{7}{}_{12} \neq 0$, there is always an O5-plane along the three-cycle parametrised by directions 347, sourcing charge on the dual four-cycle 1256. It is therefore convenient to impose an orientifold involution that generates these O5-planes. This is realised by $\sigma_\alpha$, which generates O5-planes at $x^{1,2,5,6} = \{0,1\}$ as can be seen from the lattice \eqref{eq:nil_lattice},  and all the other intersecting O5-plane images are then generated by the orbifold actions. One can work with the standard $\mathbb{Z}_2^3$ orbifold, but that one generates O5-planes along all seven three-cycles, and hence one should introduce D5-branes such that there is net vanishing charge along particular four-cycles. Another option is to work with $\mathbb{Z}_2^3$ orbifolds that include shifts which reduce the orbifold and orientifold fixed points. The orbifolds act on the entire lattice, and one can introduce shifts in the coordinates as long as they leave the coframe one-forms invariant, see ref.~\cite{Andriolo:2018yrz}. A safe option is to only include shifts in directions given by the centre of the nilmanifolds, here 5, 6 and 7, as they do not introduce twists in other directions. Since the periodicity of the coordinates is now 2 instead of one, one should change the non-vanishing shifts from 1/2 to 1 instead. The shifts that do the job and the orbifold actions that have fixed points are summarised in Table \ref{tab:shifts}.
\begin{table}[ht]
    \centering
    \begin{tabular}{|c||c|c|c|c|c|c|c|c|c||c|} \hline
        Nilmanifold & $a_5$ & $a_6$ & $a_7$ & $b_5$ & $b_6$ & $b_7$ & $c_5$ & $c_6$ & $c_7$ & Orbifold f.p. \\ \hline
        $\mathfrak{n}_2$ & {} & {} & \cellcolor{black} & {} & {} & {} & \cellcolor{black} & \cellcolor{black} & {} & $\hat \beta$ \\\hline
        $\mathfrak{n}_4$ & {} & \cellcolor{black} & \cellcolor{black} & {} & {} & {} & \cellcolor{black} & \cellcolor{black} & {} & $\hat \beta$, $\hat \alpha \hat \beta \hat \gamma$\\\hline
        $\mathfrak{n}_6$ & \cellcolor{black} & \cellcolor{black} & \cellcolor{black} & \cellcolor{black} & \cellcolor{black} & \cellcolor{black} & \cellcolor{black} & \cellcolor{black} & {} & $\hat \alpha\hat \beta$, $\hat \beta \hat \gamma$, $\hat \alpha \hat \gamma$\\\hline
        37B$_1$ & {} & {} & \cellcolor{black} & \cellcolor{black} & {} & \cellcolor{black} & \cellcolor{black} & {} & {} & $\hat \gamma$, $\hat \beta \hat \gamma$, $\hat \alpha \hat \beta \hat \gamma$\\\hline
        37C & {} & {} & \cellcolor{black} & \cellcolor{black} & \cellcolor{black} & \cellcolor{black} & {} & \cellcolor{black} & {} & $\hat \alpha \hat \gamma$, $\hat \alpha \hat \beta \hat \gamma$ \\\hline
        37D$_1$ & {} & \cellcolor{black} & \cellcolor{black} & {} & \cellcolor{black} & \cellcolor{black} & {} & \cellcolor{black} & {} & $\hat \alpha \hat \beta$, $\hat \beta \hat \gamma$, $\hat \alpha \hat \gamma$\\\hline
    \end{tabular}
    \caption{Shifts for the $\mathbb{Z}_2^3$ orbifold. A black box corresponds to a shift of 1, whereas a white box corresponds to no shift. The last column summarises which orbifold actions have fixed points.}
    \label{tab:shifts}
\end{table}
Before turning to the specific well-controlled scale-separated solutions, it is useful to further make some general remarks.
For instance, the supersymmetry equations are solved for the 7 radii and the dilaton in terms of the fluxes. An important simplifying assumption in the analysis here is that all these scalars and all other quantities are required to scale \textit{homogeneously} under a rescaling of the fluxes. This also applies to the scalar torsion class $W_1$, which sets the AdS radius.
Since $W_1$ in \eqref{eq:general_W1_nil} involves a sum, homogeneity requires all terms to scale the same. The structure constants are fixed to be the same (up to signs) by the $F_3$-tadpole, which forces specific scaling relations between the radii. For instance, for nilmanifold $\mathfrak{n}_4$, it forces $L_6/L_3 \sim L_7/L_2$, with a fixed proportionality constant, which in turn, by solving the supersymmetry equations~\eqref{eq:F3BPS} and~\eqref{eq:F7BPS}, implies that some of the fluxes scale similarly to infinity.
Indeed, there seems to be a general rule for how the parametric scalings of the radii with the RR fluxes work. This is not limited to the setups here but applies to various other flux compactifications in different dimensions as well.
If the potential at the vacuum is supposed to scale homogeneously under flux rescalings, then so must the superpotential $P$. This means that all the terms must scale in the same way, and so must the terms involving the RR fluxes in \eqref{eq:WK_pot}, i.e. $ \star \Phi \wedge F_3 \sim F_7$. From eq.~\eqref{eq:Phi_and_starPhi}, one can infer that this implies that
\begin{equation}
\label{eq:scalings_radii}
    L_i \sim \left|f_7\frac{\prod_{j<k}F_{ijk}^2}{\prod_{l<m<n} F_{lmn}}\right|^{1/4}\,.
\end{equation}
Then, if one requires $W_1$ to scale homogeneously, one finds that 
\begin{equation}
    W_1 \sim f_7^{-1/4}  \left| f^{7}{}_{12} F_{347} F_{567}\right| \prod_{l<m<n} |F_{lmn}|^{-1/4}, \qquad  f^{7}{}_{12} F_{347} \sim  f^a{}_{[12} F_{34]a} \quad \text{for all $a$}\,.
\end{equation}
The second condition is the same scaling relation that arises in the last term of the non-closure of $F_3$~\eqref{eq:dF3_nilmanifolds}, and hence is consistent with the fact that all non-closed fluxes contributing to the 1234-component in the Bianchi identity have to scale the same.
One can also infer how the string coupling must scale from eq.~\eqref{eq:WK_pot}, as $F_7 \sim g_s^{-1} W_1 \star\Phi \wedge \Phi$. This results in 
\begin{equation}
    g_s \sim f_7^{-1} W_1 \mathcal{V}_7 \sim f_7^{3/4}\; W_1  \prod_{l<m<n} \left|F_{lmn}\right|^{-1/4}\,.
\end{equation}
Finally, note that only nilmanifold $\mathfrak{n}_2$ provides setups that are \textit{fully homogeneous} under flux rescalings, meaning that the setups are homogeneous under every independent flux rescaling separately, and all the other nilmanifolds are \textit{partially homogeneous} under flux rescalings.

In the following subsections, the well-controlled scale-separated solutions are presented, as well as a discussion on the failure of a weak string coupling limit for some other nilmanifolds.
When looking for well-controlled solutions with (partially) homogeneous scalings, it turns out that the aforementioned case 1a) applies to nilmanifolds $\mathfrak{n}_2$ and $\mathfrak{n}_4$ and case 2a) to nilmanifolds $\mathfrak{n}_6$, 37B$_1$, 37C and 37D$_1$. For all other nilmanifolds, no well-controlled solutions with (partially) homogeneous flux scalings were found.\footnote{Recently, well-controlled scale-separated solutions were found on nilmanifold $\mathfrak{n}_{10}$ using \textit{non-homogeneous} flux scalings in ref. \cite{Miao:2025rgf}.}
In the analysis, the fluxes are parametrised as follows:
\begin{align}
    &F_3 =  F_{127} e^{127} - F_{347} e^{347} - F_{567}\;e^{567} + F_{235} e^{235}+ F_{145} e^{145}+ F_{136} e^{136} - F_{246} e^{246}\,,\\
    &F_7 = -f_7\, e^{1234567}\,.
\end{align}
Notice that some minus signs are appearing in this definition of $F_3$.

\subsubsection*{Nilmanifold $\mathfrak{n}_2$}
The simplest nilmanifold to consider is named $\mathfrak{n}_2$ in refs.~\cite{MR2811660,MR3739330,MR4626831,MR4789076}, which can be decomposed as a product of a flat four-torus and a 3d nilmanifold arising from the Heisenberg algebra, as it has the following Maurer-Cartan equation:
\begin{equation}
    \dd e^7 = f^{7}{}_{12} \, e^{12}\,, \qquad  \dd e^{1,2,3,4,5,6} =0\,.
\end{equation}
For this nilmanifold, the torsion class $W_1$ is given as follows (with $\omega \equiv f^{7}{}_{12}$):
\begin{equation}
\label{eq:n2_W1}
    W_1 = \frac{2\omega}{7} \frac{L_7}{L_1 L_2}\,.
\end{equation}
The supersymmetry equations \eqref{eq:F3BPS} and \eqref{eq:F7BPS} fix the radii and string coupling to be the following: 
\begin{align}
\label{eq:n2_Lisol}
     &L_i = l_i \left|f_7\frac{\prod_{j<k}F_{ijk}^2}{\prod_{l<m<n} F_{lmn}}\right|^{1/4}\,, \qquad l_i = \{\sqrt{2}, \sqrt{2},1,1,1,1,1/\sqrt{2}\}\,,\\
\label{eq:n2_gssol}
     &g_s = \frac{1}{6}\omega  (F_{347}F_{567}) f_7^{1/2}\prod_{l<m<n} F_{lmn}^{-1/2}\,.
\end{align}
Note that this solution is special in the sense that $W_1$ consists of a single term, and as a consequence, the supersymmetry equations \eqref{eq:F3BPS} and \eqref{eq:F7BPS} are fully homogeneous under independent rescalings of all the fluxes.
These solutions are also solutions of the equations of motion, provided that the non-trivial Bianchi identity for $F_3$ is satisfied. The exterior derivative on $F_3$ gives
\begin{equation}
    \dd F_3 = - (\omega F_{567})\; e^{1256} - (\omega F_{347})\; e^{1234}\,,
\end{equation}
which is to be cancelled by (smeared) O5-plane charge. Since the O5-plane images from the orbifold generate the same O5-charge along different directions, it implies that $F_{347}= F_{567}$ for a consistent background. Moreover, the O5-plane charge cannot be tuned to infinity, and since the RR three-form flux and structure constants are quantised, they cannot be tuned to infinity either.
Even though these are fixed, the seven-form flux and all the other 5 three-form fluxes are independently unbounded. In the large three-form flux limit, all radii become smaller than the AdS radius, achieving scale separation.
Indeed, using, eqns.~\eqref{eq:W1isMu}, \eqref{eq:n2_W1} and \eqref{eq:n2_Lisol}, one finds that 
\begin{equation}
\label{eq:n2_scalesep}
    \frac{L_i^2}{L_\mathrm{AdS}^2} \sim \frac{1}{\prod_{k<l<m, | i\notin \{k,l,m\}} F_{klm}^{(\mathrm{ub})}}\,,
\end{equation}
where the subscript $(\mathrm{ub})$ indicates one should only include unbounded three-form fluxes.
For obtaining large radii in string units and weak string coupling, one must put extra conditions in how the large flux limit must be engineered. The simplest example of such a limit is realised by taking $F_{235}\sim F_{145}\sim F_{136} \sim F_{246}$. As can be seen from eq.~\eqref{eq:n2_Lisol}, $L_1$ and $L_2$ behave parametrically the same in this limit, as do $L_3$, $L_4$, $L_5$ and $L_6$. This simplifies eq.~\eqref{eq:n2_scalesep} to
\begin{equation}
    \frac{L_{1,2}^2}{L_\mathrm{AdS}^2} \sim \frac{1}{F_{235}^2}\,, \qquad \frac{L_{3,4,5,6}^2}{L_\mathrm{AdS}^2}\sim \frac{1}{F_{127}F_{235}^2}\,,\qquad \frac{L_{7}^2}{L_\mathrm{AdS}^2} \sim \frac{1}{F_{235}^4}\,.
\end{equation}
Additionally, by investigating eqns.~\eqref{eq:n2_Lisol}-\eqref{eq:n2_gssol}, one obtains large radii in string units and weak string coupling under the condition that
\begin{equation}
   F_{127}\ll f_7\,, \qquad f_7 F_{127}^{-1} \ll F_{235}^4 \ll f_7 F_{127}\,,
\end{equation}
where $f_7$, $F_{127}$ and $F_{235}$ are large.\footnote{One can reparametrise these bounds in terms of scaling relations, i.e. $f_7\sim N^a$, $F_{127} \sim N^c$, $F_{235} \sim N^b$. The constraints on the fluxes then translate into $N\gg 1$,  $a>c$ and $a-c < 4b < a+c$.}\footnote{
In this limit, it can be checked that the central charge of the putative holographic dual field theory behaves as follows:
\begin{equation}
    c = \frac{2}{3} \Mpl L_\mathrm{AdS} \sim f_7 F_{127} F_{235}^4 \implies f_7^2 \ll c \ll f_7^2 F_{127}^2\,.
\end{equation}}
The masses can be computed from the potential \eqref{eq:V_pot}, and the corresponding conformal dimensions are computed to be integer:
\begin{equation}
    m^2 L_\mathrm{AdS}^2 =\{120, 8,8,8,8,8,8,8\}, \qquad \Delta = \{12, 4,4,4,4,4,4,4\}\,.
\end{equation}

\subsubsection*{Nilmanifold $\mathfrak{n}_4$}
The next nilmanifold that provides scale-separated solutions in the well-controlled regime is called $\mathfrak{n}_4$, for which the Maurer-Cartan equations are the following:
\begin{equation}
    \dd e^6 = f^{6}{}_{13} e^{13}, \qquad \dd e^7 = f^{7}{}_{12} e^{12}, \qquad \dd e^{1,2,3,4,5}=0\,,
\end{equation}
Choosing now $\omega = f^{6}{}_{13} = f^{7}{}_{12}$, the scalar torsion class becomes
\begin{equation}
\label{eq:n4_W1}
    W_1 =\frac{2\omega}{7} \left( \frac{L_6}{L_1L_3}+ \frac{L_7}{L_1L_2} \right) = \frac{2\omega}{7 }\frac{ L_2 L_6 + L_3 L_7}{L_1 L_2 L_3}\,,
\end{equation}
and the exterior derivative on $F_3$ becomes
\begin{equation}
    \dd F_3 = - (\omega F_{567}) \left( e^{1256} -e^{1357}\right) - (\omega [F_{347}-F_{246}])\; e^{1234}\,,
\end{equation}
This has to be cancelled by appropriate O5-charges, which in turn implies that $\omega$ and $F_{567}$ cannot be tuned to infinity, but additionally one needs $F_{347} - F_{246}$ also to be fixed, or vanish entirely which is prevented by supersymmetry equations \eqref{eq:F3BPS} and \eqref{eq:F7BPS}.
To solve for the latter, it is useful to reparametrise one of the scalars as 
\begin{equation}
\label{eq:n4_moduli_Ansatz}
    L_7 = \rho \frac{L_2}{L_3} L_6\,,
\end{equation}
such that the torsion class $W_1 = 2 \omega (1+\rho)L_6/(7L_1 L_3)$, and homogeneous scalings require $\rho$ not to scale. For this, the solutions are again found to be
\begin{align}
\label{eq:n4_Lisol}
    & L_i = l_i \left|f_7\frac{\prod_{j,k}F_{ijk}^2}{\prod_{l,m,n} F_{lmn}}\right|^{1/4}_{F_{246} \to F_{347}}\,,\\
    & l_i^4 = \left\{2 \rho \left(\frac{2\rho-1}{\rho+1}\right)^2,\;  \frac{2}{\rho},\; 2\rho,\;  \frac{2}{\rho} \left(\frac{\rho+1}{2\rho-1}\right)^2,\; \frac{\rho}{2}\left(\frac{2\rho-1}{\rho+1}\right)^2, \; \frac{1}{2\rho}, \;  \frac{\rho}{2} \right\}\,,\\
\label{eq:n4_gssol}
    & g_s = \frac{1+\rho}{3\sqrt{2\rho}}\omega  F_{567}|F_{347}F_{246}|^{1/2} f_7^{1/2}\prod_{l,m,n} |F_{lmn}|^{-1/2}, \quad F_{246} = \frac{\rho-2}{\rho(2\rho -1)}F_{347}\,.
\end{align}
Note that the subscript $F_{246} \to F_{347}$ means that one should replace $F_{246}$ by  $F_{347}$ in that expression.
Since $F_{347}-F_{246}>0$ for all $\rho$, one needs O5-planes transverse to the 1234 directions, implying both $F_{347}$ and $F_{246}$ are fixed. In total, three sets of intersecting O5-planes are necessary (transverse to the four-cycles in the 1256-, 3456- and 1234-directions, as summarised in Table \ref{tab:nilmanifolds}), which can be generated by the orientifold involution $\sigma_{\hat{\alpha}}$ and appropriate shifts in the orbifold, the latter being summarised in Table \ref{tab:shifts}. 
However, since all intersecting orientifold images come in the same numbers, i.e. $2^4 = 16$, and because O5-planes have the same charge as D5-branes, one needs to make sure that $16 = \omega F_{567} = \omega (F_{347}-F_{246})$.
We have that 
\begin{equation}
\label{eq:tadpole_constraint_n4}
    F_{347}-F_{246} = F_{347}\left(1 - \frac{\rho-2}{\rho( 2\rho-1)}\right) = F_{567}\,.
\end{equation}
There are only a finite number of positive integer solutions. Since both $F_{567}$ and $\omega$ are quantised, there are only 5 positive options for $F_{567}$ as $N_\mathrm{O5}=16 = 1\cdot 16 = 2 \cdot 8 = 4 \cdot 4 = 8 \cdot 2 = 16 \cdot 1$. When $F_{567}=n$, there are $n$ options for $F_{347}=1,...,n$ and corresponding positive real $\rho$, leading to a total of $1+2+4 +8 +16 = 31$ different configurations, but only 16 distinct values for $\rho$ as the solutions to eq.~\eqref{eq:tadpole_constraint_n4} with $F_{567}/F_{347}=16/n$ and $n=1,...,16$.

Finally, more fluxes are restricted than for nilmanifold $\mathfrak{n}_2$, but again all radii are small in AdS units in the large three-form flux limit. One can still engineer an infinite flux limit in which the radii are large in string units and weak string coupling are achieved. By fixing $\rho$ and then taking the limit in which all remaining unbounded three-form fluxes are large and scale the same, $F_{235} \sim F_{127} \sim  F_{145} \sim F_{136}$, scale-separation is realised as advertised, as one sees from eqns.~\eqref{eq:n4_W1} and\eqref{eq:n4_Lisol} that
\begin{equation}
    \frac{L_1^2}{L_\mathrm{AdS}^2}\sim \frac{1}{F_{235}}\,, \qquad \frac{L_{2,3,5}^2}{L_\mathrm{AdS}^2}\sim \frac{1}{F_{235}^2}\,, \qquad \frac{L_{4,6,7}^2}{L_\mathrm{AdS}^2}\sim \frac{1}{F_{235}^3}\,.
\end{equation}
As can be seen from \eqref{eq:n4_Lisol}-\eqref{eq:n4_gssol}, the well-controlled regime with weak string coupling and large radii in string units is obtained when
\begin{equation}
\label{eq:n4_scale_sep}
    f_7 \ll F_{235}^4 \ll f_7^2 \iff F_{235}^2 \ll f_7 \ll F_{235}^4\,,
\end{equation}
and both $F_{235}$ and $f_7$ are large.
The masses and conformal dimensions depend explicitly on $\rho$, and they do not lead to integer conformal dimensions. More details on the mass spectra can be found in Appendix \ref{app:masses}.

\subsubsection*{Nilmanifold $\mathfrak{n}_6$}
Consider now the last decomposable nilmanifold that gives rise to scale-separated and well-controlled solutions, $\mathfrak{n}_6$. This space can be regarded as a direct product of a circle and the 6d Iwasawa manifold.
In this case, the Maurer-Cartan equations are 
\begin{equation}
    \dd e^6 = f^{6}{}_{13} e^{13} + f^{6}{}_{24}e^{24}\,, \qquad \dd e^7 = f^{7}{}_{12} e^{12}+ f^{7}{}_{34} e^{34}\,, \qquad \qquad \dd e^{1,2,3,4,7}=0\,.
\end{equation}
By defining $\omega \equiv f^{7}{}_{12} = f^{7}{}_{34}=f^{6}{}_{13}= - f^{6}{}_{24}$, one sees that the torsion class in the trivial representation becomes
\begin{equation}
    W_1 = \frac{2\omega}{7} \frac{L_1 L_3 L_6 + L_2 L_4 L_6 + L_1 L_2 L_7 + L_3 L_4 L_7}{L_1 L_2 L_3 L_4}\,,
\end{equation}
and the non-closure of $F_3$ is captured by 
\begin{equation}
\label{eq:dF3_n6}
    \dd F_3 = -(\omega F_{567}) \left(e^{1256}+e^{3456}+e^{2457}-e^{1357} \right)  -\omega(F_{127}- F_{347}+F_{136}+F_{246})  e^{1234}\,.
\end{equation}
As for all other nilmanifolds, $\omega$ and $F_{567}$ are fixed by tadpole cancellation with O5-planes, and the sum $(F_{127}-F_{347}+F_{136}+F_{246})$ as well, unless there is a way for it to vanish. Additionally, it turns out that one cannot get large radii and weak string coupling at the same time, unless this combination vanishes. To find specific solutions of the supersymmetry equations \eqref{eq:F3BPS} and \eqref{eq:F7BPS}, one can start from the following simplifying Ansatz: 
\begin{equation}
\label{eq:moduli_Ansatz_n6}
    L_7 = \rho L_6, \qquad L_3 = \tau L_2, \qquad L_4 = \tau L_1\,,
\end{equation}
with which the following solution is obtained:
\begin{align}
\label{eq:n6_Lisol}
     &L_1 = 2^{1/4}\tau^{-1/2}\left(\frac{f_7 F_{145}}{F_{235}F_{567}} \right)^{1/4}\,, \quad L_2 = 2^{1/4}\tau^{-1/2}\left(\frac{f_7 F_{235}}{F_{145}F_{567}} \right)^{1/4}\,,\\
     &L_5 = 2^{-1/4}\rho^{-1/2}\frac{\rho-\tau + \rho \tau^2}{\rho + 2 \tau + \rho \tau^2}\frac{1}{F_{246}}\left(f_7 F_{145}F_{235}F_{567} \right)^{1/4}\,, \quad L_6 = 2^{-1/4}\rho^{-1/2}\left(\frac{f_7 F_{567}}{F_{145}F_{235}} \right)^{1/4},\\
\label{eq:n6_gssol}
     &g_s = \frac{\omega (\rho-\tau + \rho \tau^2)}{3 \sqrt{2} \rho \tau}\frac{1}{F_{246}}\left(\frac{f_7 F_{567}}{F_{145}F_{235}}\right)^{1/2}\,, \qquad F_{136}= F_{246}\,,\\
\label{eq:n6_fluxsolutions}
     &F_{127}=\frac{\rho  \left(\rho  \left(\tau ^2-2\right)+2 \tau
   \right)}{\tau  \left(\rho  \tau ^2+\rho -\tau
   \right)}F_{246}, \quad F_{347} = \frac{\rho  \tau  \left(\rho  \left(2 \tau
   ^2-1\right)-2 \tau \right)}{\rho  \tau ^2+\rho -\tau
   }F_{246} \,.
\end{align}
The solutions require also that all $F_{235}, F_{145}, f_7, \rho,\tau$ are positive and that $F_{246}$ and $(\rho-\tau+\rho\tau^2)$ have the same sign.
The last term in eq.~\eqref{eq:dF3_n6} vanishes if
\begin{equation}
\label{eq:n6_rhotau_sol}
    \rho = \tau \frac{1 \pm \sqrt{3}\tau+ \tau^2}{1-\tau^2 + \tau^4}\,.
\end{equation}
Note however that the only solutions to this equation that are allowed are such that the combinations in \eqref{eq:n6_fluxsolutions} are rational. This happens for instance when $\tau \in \sqrt{3}\:\mathbb{Q}$, e.g.~for $(\tau, \rho) = \sqrt{3}\left((1,(4\pm 3)/7\right)$, but also for $(\tau,\rho) = (1, 2 \pm \sqrt{3})$. Additionally, $F_{246}$ is negative (positive) for the $-$ ($+$) branch of condition \eqref{eq:n6_rhotau_sol}.
When eq.~\eqref{eq:n6_rhotau_sol} is satisfied, only four sets of intersecting O5-planes are necessary, summarised in Table \ref{tab:nilmanifolds}, which are realised by the orientifold involution and orbifold with the shifts found in Table \ref{tab:shifts}. 
As usual, scale separation with the radii is achieved in the large three-form flux limit, and weak string coupling and large radii in string units can be achieved by taking a proper limit. Indeed, fixing $\tau$ and $\rho$ through eq.~\eqref{eq:n6_rhotau_sol}, and taking a limit where $F_{235}\sim F_{145}\sim  F_{246}$, scale separation is achieved as
\begin{equation}
    \frac{L_{1,2,3,4}^2}{L_\mathrm{AdS}^2}\sim \frac{1}{F_{235}}, \qquad \frac{L_{5,6,7}^2}{L_\mathrm{AdS}^2}\sim \frac{1}{F_{235}^2}\,,
\end{equation}
on the condition that $F_{235}$ is large. Additionally, weak coupling and large radii in string units are guaranteed by \eqref{eq:n6_Lisol}-\eqref{eq:n6_gssol} when in addition $f_7$ is large and
\begin{equation}
\label{eq:flux_limit_n6}
    f_7 \ll F_{235}^4 \ll f_7^2 \iff F_{235}^2 \ll f_7 \ll F_{235}^4 \,.
\end{equation}
However, the masses and conformal dimensions depend on $\tau$, and they are not all integers. Moreover, only for a small window of one of the two branches of \eqref{eq:n6_rhotau_sol} are all masses squared positive, and in the large $\tau$-limit, some of the scalars become very light in AdS units. More details on this can be found in appendix \ref{app:masses}.
\subsubsection*{Nilmanifold 37B$_1$}
Having exhausted well-controlled scale-separated vacua on decomposable nilmanifolds, one should turn to the indecomposable ones. Of all of those, only 37B$_1$, 37C and 37D$_1$ can give rise to scale-separated vacua in the well-controlled regime.
Let us start with 37B$_1$. The algebra here is
\begin{equation}
    \dd e^5 = f^5{}_{14} e^{14}, \qquad \dd e^6 = f^6{}_{13} e^{13} + f^6{}_{24} e^{24}, \qquad \dd e^7 = f^7{}_{12} e^{12} + f^7{}_{34} e^{34}\,, \dd e^{1,2,3,4}=0\,.
\end{equation}
After taking $\omega \equiv f^{5}_{14}=f^{6}_{13}=f^{7}_{12}=f^{7}_{34} = - f^{6}_{24}$, the torsion class $W_1$ becomes
\begin{equation}
\label{eq:37B1_W1}
    W_1 = \frac{2\omega}{7}\frac{L_2 L_3 L_5 + L_1 L_3 L_6 + L_2 L_4 L_6 + L_1 L_2 L_7 + L_3 L_4 L_7}{L_1 L_2 L_3 L_4}\,,
\end{equation}
and the non-closure of $F_3$ becomes
\begin{align}
\label{eq:dF3_37B1}
\begin{split}
        \dd F_3 =& - (\omega F_{567})\left(e^{1256} + e^{3456} + e^{1467} + e^{2457} - e^{1357} \right)\\
        &+ \omega (F_{127}-F_{347}+ F_{145} + F_{136} + F_{246})e^{1234}\,.
\end{split}
\end{align}
If one wants $W_1$ to scale homogeneously with the fluxes, then one must require that \linebreak $L_5\sim L_6 \sim L_7$ and $L_1 \sim L_2 \sim L_3 \sim L_4$. There are many solutions of this kind, and for this section, the following Ansatz is chosen
\begin{equation}
\label{eq:moduli_Ansatz_37B1}
        L_3 = L_1, \qquad L_4 = L_2 = \tau L_1, \qquad L_6 = \tau L_5, \qquad L_7 = \rho L_5\,.
\end{equation}
For this Ansatz, the solution of the supersymmetry equations \eqref{eq:F3BPS} and \eqref{eq:F7BPS} fix the remaining radii and string coupling as follows:
\begin{align}
\label{eq:37B1_Lisol}
    &L_1 = 2^{1/4} \tau^{-1/2}\left(\frac{f_7}{F_{567}}\right)^{1/4}\,, \qquad L_5 = 2^{-1/4}\sqrt{
    \frac{-1+2\rho+\tau^2}{\rho\tau(2+2\rho+\tau^2)}} \left(\frac{f_7 F_{567}}{F_{235}^2}\right)^{1/4}\,,\\
\label{eq:37B1_gssol}
    &g_s = \frac{\omega}{3 \sqrt{2}} 
    \frac{\tau(-1+2\rho+\tau^2)^2}{\rho\tau(2+2\rho+\tau^2)} \left(\frac{f_7 F_{567}}{F_{235}^4}\right)^{1/2}, \qquad
    F_{145}=\frac{2+2\rho +\tau^2}{-1+2\rho +\tau^2}F_{235}\,,\\
   &F_{246}=\frac{2\tau^2 (1+\rho-\tau^2)}{-1+2\rho+\tau^2} F_{235}, \qquad  F_{127} = -F_{347} = -\frac{\rho(-2+\rho-\tau^2)}{-1+2\rho+\tau^2} F_{235}\,,\\
   &F_{136}=F_{235}\,.
\end{align}
Note that $f_7$, $F_{567}$, $F_{235}$, $\rho$, $\tau$ and $\omega$ are all positive 
The 1234-component in the non-closure of $F_3$ \eqref{eq:dF3_37B1} vanishes on the condition that
\begin{equation}
\label{eq:37B1_rhotau_sol}
    \rho = 2 + \tau^2 \pm \sqrt{3}\sqrt{1+2\tau^2}\,,
\end{equation}
which must be solved such that the relations between the fluxes are rational, otherwise it is inconsistent with flux quantisation. A nice example is $(\tau,\rho) = (6,1)$.
For all these cases, there are 5 intersecting sets of O5-planes required to cancel the $F_3$-tadpole from \eqref{eq:dF3_37B1}, which are summarised in Table \ref{tab:nilmanifolds} and the orbifolds shifts that generate these intersecting images are displayed in Table \ref{tab:shifts}.

Due to tadpole cancellation with O5-planes, the parameter $\omega$ and three-form flux $F_{567}$ are fixed by the O5-plane charge and the only independent unbounded fluxes are $f_7$ and $F_{235}$. 
Upon fixing $\tau$ (and $\rho$), scale separation with all radii is achieved when this three-form flux is large: from $L_\mathrm{AdS}\sim 1/W_1$ and \eqref{eq:37B1_W1}, \eqref{eq:moduli_Ansatz_37B1}, \eqref{eq:37B1_Lisol}, one sees that
\begin{equation}
    \frac{L_{1,2,3,4}^2}{L_\mathrm{AdS}^2} \sim \frac{L_5^2}{L_1^2} \sim \frac{1}{F_{235}}, \qquad \frac{L_{5,6,7}^2}{L_\mathrm{AdS}^2} \sim \frac{L_5^4}{L_1^4} \sim \frac{1}{F_{235}^2}\,.
\end{equation}
Additionally, large radii in string units and weak string coupling are realised when
\begin{equation}
\label{eq:flux_limit_37B1}
    f_{7} \ll F_{235}^4 \ll f_7^2 \iff F_{235}^2 \ll f_7 \ll F_{235}^4\,, 
\end{equation}
as can be seen from eqns.~\eqref{eq:37B1_Lisol}-\eqref{eq:37B1_gssol} where $F_{235}$ and $f_7$ are large.
Moreover, just as for nilmanifold $\mathfrak{n}_6$, the masses of the eight scalars and their conformal dimensions depend on $\tau$ (and $\rho$) and are non-integer. 
For generic values of $\tau$, there is a scalar with negative masses squared (above the BF bound), and the large $\tau$-limit leads to very light scalars with respect to the AdS scale. For more details, see appendix \ref{app:masses}.

\subsubsection*{Nilmanifold 37C}
Nilmanifold 37C is parametrised by the following Maurer-Cartan equations:
\begin{equation}
    \dd e^5 = f^{5}{}_{23} e^{23}, \qquad \dd e^6 = f^{6}{}_{24} e^{24}, \qquad \dd e^7 = f^{7}{}_{12} e^{12} + f^{7}{}_{34} e^{34}\,, \qquad \dd e^{1,2,3,4}=0\,.
\end{equation}
After choosing $\omega \equiv f^{5}_{23}=f^{7}_{12}=f^{7}_{34} = - f^{6}_{24}$, the torsion class $W_1$ becomes
\begin{equation}
\label{37C_W1}
    W_1 = \frac{2 \omega}{7}\frac{L_1 L_4 L_5 + L_1 L_3 L_6 + L_1 L_2 L_7 + L_3 L_4 L_7}{L_1 L_2 L_3 L_4}\,,
\end{equation}
and the non-closure of $F_3$ is given as follows:
\begin{equation}
\label{eq:dF3_37C}
    \dd F_3 = - (\omega F_{567})\left(e^{1256} + e^{3456} + e^{2367} + e^{2457}\right) + \omega (F_{127}-F_{347}+ F_{145} + F_{136})e^{1234}
\end{equation}
Also for this nilmanifold, homogeneous scalings are obtained when $L_5 \sim L_6 \sim L_7$ and \linebreak $L_1 \sim L_2 \sim L_3 \sim L_4$. Taking again the simplifying Ansatz $L_1 = L_2 = L_3 = L_4$, and 
\begin{equation}
\label{eq:moduli_Ansatz_37C}
    L_6 = \tau L_5 , \quad L_7 = \rho L_5\,,
\end{equation}
one finds that the supersymmetry conditions are solved by the following quantities
\begin{align}
\label{eq:37C_Lisol}
    &L_1  =  2^{1/4}\left(\frac{f_7}{F_{567}}\right)^{1/4}\,, \qquad L_5  = 2^{-1/4}(\rho \tau)^{-1/2}
   \left(\frac{F_{567}f_7}{F_{235}^2}\right)^{1/4}\,,\\
\label{eq:37C_gssol}
   &g_s  = 
   \frac{( \tau+2  \rho +1)}{3 \sqrt{2}  \tau  \rho}\omega \left(\frac{f_7 F_{567}}{F_{235}^4}\right)^{1/2}\,, \qquad   F_{127} = -F_{347} = -\frac{\rho (- \tau+ \rho -1)}{ \tau+2  \rho+1} F_{235}\,,\\
   &F_{145} =  \frac{( \tau+2  \rho -2)}{ \tau+2  \rho+1}F_{235}\,, \qquad 
   F_{136} =  -\frac{\tau (2  \tau-2  \rho-1)}{ \tau+2  \rho+1} F_{235}\,, \qquad F_{246} =   \tau F_{235}\,.
\end{align}
Note that $f_7$, $F_{567}$, $F_{235}$, $F_{246}$, $\rho$, $\tau$ and $\omega$ are all positive.
Also here, the interesting case is when the last term of the non-closure of $F_3$ \eqref{eq:dF3_37C} vanishes, which happens when
\begin{equation}
\label{eq:37C_rhotau_sol}
   \rho = 1 + \tau \pm \sqrt{3\tau}\,.
\end{equation}
As before, it is important that the solutions to this equation must lead to rational relations between the fluxes due to flux quantisation, and hence, not all solutions are allowed. However, the simple solutions $(\tau , \rho) = (3,1)$ and $(3,7)$ do so.
For all these cases, there are 4 intersecting sets of O5-planes required to cancel the $F_3$-tadpole from \eqref{eq:dF3_37B1}. The necessary O5-planes can be found in Table~\ref{tab:nilmanifolds}, and the orbifold shifts that generate these intersecting images are displayed in Table~\ref{tab:shifts}.
Very similarly to the cases above, scale separation with the radii is achieved whenever one fixes $\tau$ (and $\rho$ through eq.~\eqref{eq:37C_rhotau_sol}) and takes $F_{235}$ large. This can be seen from eqns.~\eqref{eq:37B1_W1} (with $L_\mathrm{AdS} \sim 1/W_1$), \eqref{eq:moduli_Ansatz_37C} and \eqref{eq:37C_Lisol}, as
\begin{equation}
    \frac{L_{1,2,3,4}^2}{L_\mathrm{AdS}^2} \sim \frac{L_5^2}{L_1^2} \sim \frac{1}{F_{235}}, \qquad \frac{L_{5,6,7}^2}{L_\mathrm{AdS}^2} \sim \frac{L_5^4}{L_1^4} \sim \frac{1}{F_{235}^2}\,.
\end{equation}
Similarly, one obtains large radii and weak string coupling from eqns.~\eqref{eq:37C_Lisol}-\eqref{eq:37C_gssol} by choosing
\begin{equation}
\label{eq:flux_limit_37C}
    f_7 \ll F_{235}^4 \ll f_7^2 \iff F_{235}^2 \ll f_7 \ll F_{235}^4\,,
\end{equation}
where $F_{235}$ and $f_7$ are large.
The masses of the eight scalars and their conformal dimensions depend on $\tau$ (and $\rho$) and are not integers again. For generic values of $\tau$, there are modes with negative masses squared (above the BF bound), and the large $\tau$-limit leads to light scalars, which is described in detail in appendix \ref{app:masses}.
\subsubsection*{Nilmanifold 37D$_1$}
For this last nilmanifold, all structure constants are turned on, and the Maurer-Cartan equations look as follows:
\begin{equation}
    \dd e^{5} = f^{5}{}_{23} e^{23} + f^{5}{}_{14} e^{14}, \quad \dd e^{6} = f^{6}{}_{13} e^{13} + f^{6}{}_{24} e^{24}, \quad \dd e^{7} = f^{7}{}_{12} e^{12} + f^{7}{}_{34} e^{34}, \quad \dd e^{1,2,3,4} = 0\,.
\end{equation}
One needs to take $\omega \equiv f^{5}{}_{14} = f^{5}{}_{23} = f^{6}{}_{13} = -f^{6}{}_{24} =f^{7}{}_{12}=f^{7}{}_{34}$, for which the scalar torsion class becomes
\begin{equation}
\label{eq:37D1_W1}
    W_1 = \frac{2\omega}{7}\frac{L_1 L_4 L_5 + L_2 L_3 L_5 + L_1 L_3 L_6 + L_2 L_4 L_6 + L_1 L_2 L_7 + L_3 L_4 L_7}{L_1 L_2 L_3 L_4}\,,
\end{equation}
and the non-closure of $F_3$ is now given by:
\begin{align}
\begin{split}
\label{eq:dF3_37D1}
    \dd F_3 =& - (\omega F_{567})\left(e^{1256} + e^{3456} + e^{1467} + e^{2457} + e^{2367} - e^{1357} \right)\\
    &+ \omega (F_{127}-F_{347}+ F_{145} + F_{136} + F_{235} + F_{246})e^{1234}\,.
\end{split}
\end{align}
Similar to previous cases, one needs to require $L_1 \sim L_2 \sim L_3 \sim L_4$ and $L_5 \sim L_6 \sim L_7$ such that quantities like $W_1$ scale homogeneously. This time, another Ansatz is more useful:
\begin{equation}
\label{eq:moduli_Ansatz_37D1}
    L_3 = L_1, \qquad L_4 = L_2 = \tau L_1, \qquad L_6 = \tau L_5, \qquad L_7 = \rho L_5.
\end{equation}
With those, the solutions of the supersymmetry equations \eqref{eq:F3BPS} and \eqref{eq:F7BPS} become:
\begin{align}
\label{eq:37D1_Lisol}
    &L_1 =  2^{1/4} \tau^{-1/2} \left(\frac{ f_{7}}{ F_{567}}\right)^{1/4}\,, \qquad 
   L_5 =  2^{-1/4} \sqrt{\frac{2 \rho + \tau^2}{\rho \tau(3 + 2 \rho + \tau^2)}}\left(\frac{F_{567}f_7}{F_{235}^2} \right)^{1/4}\,,\\
\label{eq:37D1_gssol}
   & g_s =\frac{\omega}{3\sqrt{2}} \left(\frac{(2 \rho + \tau^2)^2}{\rho \tau(3 + 2 \rho + \tau^2)}\right)\left(\frac{F_{567}f_7}{F_{235}^4} \right)^{1/2}, \qquad F_{145} = F_{136}= F_{235}\,,\\
   \label{eq:37D1_fluxsolutions}
   &F_{127} = - F_{347} = -\frac{\rho (-3 + \rho - \tau^2)}{2 \rho + \tau^2} F_{235}\,, \qquad  F_{246} =  -\frac{\tau^2 (-3 - 2 \rho + 2 \tau^2)}{2 \rho + \tau^2}F_{235}\,.
\end{align}
The quantities $f_7$, $F_{567}$, $F_{235}$, $\rho$, $\tau$ and $\omega$ are all positive.
The last term of eq.~\eqref{eq:dF3_37D1} vanishes when
\begin{equation}
\label{eq:37D1_rhotau_sol}
    \rho = 3+ \tau ^2 \pm 3 \sqrt{\tau ^2+1}\,,
\end{equation}
and the solutions of these equations must allow for rational relations between the fluxes in \eqref{eq:37D1_fluxsolutions}. This is easily arranged when $\tau^2 \in \mathbb{Z}^2-1$. The simplest solutions with $\rho,\tau>0$ are then $(\rho,\tau^2) = (12,3)$, $(2,8)$ and $(20,8)$ and so on.
If the condition \eqref{eq:37D1_rhotau_sol} is met, then there are 6 intersecting sets of O5-planes required to cancel the $F_3$-tadpole from \eqref{eq:dF3_37D1}, which are summarised in Table \ref{tab:nilmanifolds} and the orbifolds shifts that generate these intersecting images are displayed in Table \ref{tab:shifts}.
Exactly as in the previous examples, a hierarchy between the AdS scale and all the radii is obtained in the large $F_{235}$ limit by fixing $\tau$ (and $\rho$) and combining \eqref{eq:W1isMu}, \eqref{eq:37D1_W1}, \eqref{eq:moduli_Ansatz_37D1}, \eqref{eq:37D1_Lisol}:
\begin{equation}
    \frac{L_{1,2,3,4}^2}{L_\mathrm{AdS}^2} \sim \frac{L_5^2}{L_1^2} \sim \frac{1}{F_{235}}, \qquad \frac{L_{5,6,7}^2}{L_\mathrm{AdS}^2} \sim \frac{L_5^4}{L_1^4} \sim \frac{1}{F_{235}^2}\,.
\end{equation}
The string coupling is small and all radii large are large in string units as long as the unbounded three- and seven-form fluxes are large and satisfy
\begin{equation}
\label{eq:flux_limit_37D1}
    f_7 \ll F_{235}^4 \ll f_7^2 \iff F_{235}^2 \ll f_7 \ll F_{235}^4\,.
\end{equation}
This is the same condition as for all other nilmanifolds, except $\mathfrak{n}_2$.\footnote{In this limit, it can also be checked that for all these nilmanifolds, the central charge of the putative holographic dual field theory behaves as follows:
\begin{equation}
    c = \frac{2}{3}\Mpl L_\mathrm{AdS} \sim F_{235}^3 f_7 \implies F_{235}^5 \ll c \ll F_{235}^7 \quad \text{or} \quad f_7^{7/4} \ll c \ll f_7^{5/2}\,.
\end{equation}}
As for most previous nilmanifolds, the masses of the eight scalars and their conformal dimensions depend on $\tau$ (and $\rho$) and are non-integer either. For generic values of $\tau$, there are modes with negative masses squared (above the BF bound), and the large $\tau$-limit leads to light scalars, which is depicted in detail in appendix \ref{app:masses}.

\subsubsection*{No well-controlled regime for some other nilmanifolds}
For some other nilmanifolds, there does not exist a well-controlled regime while still stabilising all moduli, as the string coupling becomes large in the large radii limit.
This is illustrated most easily with nilmanifold $\mathfrak{n}_3$, which has the following Maurer-Cartan equations:
\begin{equation}
    \dd e^{7} = f^{7}{}_{12} e^{12} +  f^{7}{}_{34} e^{34}\,, \qquad \dd e^{1,2,3,4,5,6} =0\,.
\end{equation}
With $\omega \equiv f^{7}{}_{12} = f^{7}{}_{34}$, the scalar torsion class becomes
\begin{equation}
    W_1 = \frac{2\omega}{7} \left(\frac{L_7}{L_1 L_2} +\frac{L_7}{L_3 L_4} \right)\,,
\end{equation}
and the non-closure of $F_3$ is given by 
\begin{equation}
\label{eq:dF3_n3}
    \dd F_3 = -\omega F_{567}\left(e^{1256}+ e^{3456}\right) + \omega(F_{127}-F_{347})e^{1234}\,.
\end{equation}
It is instructive to look at the last term in equation \eqref{eq:dF3_n3}. Using the supersymmetry equations, it is equal to:
\begin{equation}
    F_{127}-F_{347} = -\frac{2\omega}{3 g_s}L_7^2 \frac{(L_1L_2)^2- L_1 L_2 L_3 L_4 + (L_3 L_4)^2}{L_1 L_2 L_3 L_4}\,.
\end{equation}
This cannot vanish for any non-zero values of $L_{1,2,3,4}$, and due to tadpole cancellation, \linebreak $F_{127}-F_{347}$ is fixed by O5-plane charge. The homogeneous scaling property requires $L_1 L_2 \sim L_3 L_4$ and because $\omega$ does not scale, one must impose that
\begin{equation}
    \frac{L_7^2}{g_s} \sim 1\,, 
\end{equation}
and hence $L_7$ and $g_s$ cannot be large and small at the same time, respectively. Going away from the non-homogeneous regime does not work either. Indeed, if one takes, without loss of generality, $L_1 L_2 \gg L_3 L_4$, then 
\begin{equation}
    F_{127}-F_{347} \approx -\frac{2\omega}{3 g_s}L_7^2 \frac{L_1L_2}{L_3 L_4}\,.
\end{equation}
Again, tadpole cancellation would require this not to scale, which implies that the string coupling must be large if one requires large radii.
Using (partially) homeogenously scaling fluxes, all remaining nilmanifolds suffer from a similar issue when stabilising all moduli, and therefore do not lead to well-controlled scale-separated vacua either.

\subsubsection*{Some comments on the mass spectra}
So far, the different well-controlled scale-separated solutions have been discussed, but it is also interesting to look at the mass spectra of these solutions. More details can be found in appendix \ref{app:masses}, but the important observations are summarised here.

Of all the solutions described above, nilmanifold $\mathfrak{n}_2$ provides the simplest setup. The solutions on this nilmanifold are fully homogeneous under flux rescalings, and as a consequence, the masses of the scalars, expressed in AdS units, do not depend on the fluxes. They also lead to integer conformal dimensions for the putative dual field theory operators. Although nilmanifolds $\mathfrak{n}_3$ and 17 do not lead to well-controlled (but scale-separated) solutions, they also exhibit these features.

This is no longer true for all the other solutions presented here: they are only partially homogeneous under the flux rescalings, and as a consequence, the scalar masses do depend non-trivially on the fluxes, more precisely on specific ratios involving fluxes that appear in the 1234-component of the tadpole, i.e. the ones that are not trivially unbounded. For the all the nilmanifolds  $\mathfrak{n}_4$,  $\mathfrak{n}_6$, 37B$_1$, 37C and 37D$_1$, these are given by the parameters $\rho$ and $\tau$ that were introduced the scalar Ansatze \eqref{eq:n4_moduli_Ansatz}, \eqref{eq:moduli_Ansatz_n6}, \eqref{eq:moduli_Ansatz_37B1}, \eqref{eq:moduli_Ansatz_37C} and \eqref{eq:moduli_Ansatz_37D1}, and hence the scalar masses become functions of $\tau$ (after fixing $\rho = \rho(\tau)$ through eqns.~\eqref{eq:n6_rhotau_sol}, \eqref{eq:37B1_rhotau_sol}, \eqref{eq:37C_rhotau_sol} and \eqref{eq:37D1_rhotau_sol}). 
The scalar masses are plotted as continuous functions in appendix~\ref{app:masses}, but note that only values are allowed that lead to rational ratios between the fluxes due to flux quantisation. 
Nevertheless, there are limits for $\tau$ and $\rho$ such that the masses asymptote to non-zero integer values and integer conformal dimensions. 
For $\mathfrak{n}_4$, which has no non-trivially unbounded fluxes, there is only a finite number of configurations and hence a finite number of options for $\rho$ in \eqref{eq:n4_moduli_Ansatz}. For all other nilmanifolds, i.e. $\mathfrak{n}_6$, 37B$_1$, 37C and 37D$_1$, where there are non-trivially unbounded fluxes, there do exist infinite flux limits for which some of the scalars become light in AdS units, i.e. by taking $\tau \gg 1$ first and then sending the unbounded fluxes to infinity appropriately. This is  illustrated in detail in appendix~\ref{app:masses}, and something very similar happens for a type IIA AdS$_3$ solution in ref.~\cite{Farakos:2025bwf}.
There are no masses that blow up, and in this sense the setups satisfy the AdS moduli conjecture of ref.~\cite{Gautason:2018gln} and do not provide examples of `rigid compactifications' \cite{Delgado:2025crl}.
However, it is certainly non-standard in AdS flux compactifications relying solely on classical ingredients for moduli stabilisation that the masses vary with the fluxes, that $m^2 L_\mathrm{AdS}^2$ changes sign for some fields as the fluxes vary, and that certain scalars become very light in specific flux limits.
In the context of the Swampland program, it is therefore of interest to investigate whether additional constraints might exist that would invalidate these constructions involving non-trivially unbounded fluxes.

\subsubsection*{Comments on using weak $G_2$-structures}
So far, co-closed $G_2$-structures have been considered with non-vanishing torsion class $W_{27}$. This torsion class allows for non-positive scalar curvature, which requires O5-planes to solve the Bianchi identities.
However, it is also worthwhile to contemplate setups for which $W_{27}$ vanishes. In that case, only $W_1$ survives, and the manifold is said to have a weak $G_2$-structure. Notice that $W_1$ is related to the inverse AdS radius and hence taking it to be vanishing would lead to Minkowski setups instead. So consider a type IIB background for which $W_1 \neq 0$ and $W_{27}=0$. It follows from eq.~\eqref{eq:F3smeared} that
\begin{equation}
   \dd  F_3 = g_s^{-1}\frac{W_1^2}{6} \star \Phi\,.
\end{equation}
Consequently, the Bianchi identity should be solved by an appropriate number of D5-branes wrapping three-cycles. The flux is not restricted in that case, as the number of D5-branes can be chosen at will.
However, if one wants scale separation, then one would need an anisotropic manifold for which the torsion class $W_1$ and hence the scalar curvature radius decouple from the KK radius, estimated by the three- and four-cycle volume scales. Indeed, if the manifold is isotropic, then $\mu \sim W_1 \sim 1/\mathcal{V}_7^{1/7}$, leading to no separation of scales. Furthermore, all homogeneous and smooth weak $G_2$-manifolds have been classified, and they do not allow for this property. Nevertheless, a weak $G_2$-structure manifold with these properties has recently been proposed in ref.~\cite{VanHemelryck:2024bas}, arising from a lift of a massless type IIA solution \cite{Cribiori:2021djm} to M-theory, and could therefore lead to a hierarchy of scales in these 3d compactifications as well. However, exploring this scenario in full detail requires a more complete understanding of this space, which is left for future research.

\section{Summary}
\label{sec:summary}
This paper investigates compactifications of type IIB string theory on seven-dimensional manifolds equipped with co-closed $G_2$-structures that arise as $\mathbb{Z}_2^3$ orbifolds of twisted tori and preserve minimal supersymmetry. More specifically, the geometries considered here are one solvmanifold and all nilmanifolds compatible with the orbifold. 
All the solutions require orientifolds, such that the tadpole from the $F_3$ Bianchi identity can be cancelled by O5-planes wrapping the right three-cycles. Since the amount of orientifold planes cannot be tuned in string theory, some components of the $F_3$ flux must be fixed by the tadpole, whereas others are unbounded. It is by sending the latter fluxes and the seven-form flux to large values that the well-controlled and scale-separated regime can be achieved for these vacua. 
Crucially, these backgrounds do not require O5-planes to be wrapped along all seven three-cycles. Shifts were implemented in the $\mathbb{Z}_2^3$ orbifold actions such that only the necessary intersecting orientifold images are generated. Note that in principle, the $F_3$ Bianchi identity can also be solved by wrapping O5-planes along all seven three-cycles as in ref.~\cite{Emelin:2021gzx} (realised by orbifolds without shifts), but by also including D5-branes to cancel the O5-charge along some cycles. It also leads to orbifold singularities for which it is unknown how to resolve them, even for the ordinary seven-torus. Additionally, such orbifolds also generate an O9-plane image and therefore the model resides in type I string theory instead (after adding 32 D9-branes). Working with the different orbifolds of this paper does not require the inclusion of D5-branes and keeps the setup in type IIB string theory.

On the solvmanifolds, the large flux limit leads to solutions in the well-controlled regime, i.e. with weak string coupling and large cycles in string units, and with a hierarchy between the AdS scale and the cycle volumes. However, these cycle volumes do not seem to provide good proxies for the KK scale, which is usually estimated by the lowest eigenvalue of the scalar Laplacian. Therefore it cannot be concluded that these solutions are scale-separated. The reason is that preliminary analysis shows that the lowest eigenvalue is given by the biggest length scale of the manifold ($L_7$) which, through supersymmetry, behaves parametrically the same  as the AdS radius. Note that the same issue arises in the setup of refs.~\cite{Arboleya:2024vnp,Arboleya:2025ocb} although they are non-supersymmetric.
Indeed, the solution with only one set of O5-planes can be understood as a supersymmetric cousin of one of the vacua of ref.~\cite{Arboleya:2024vnp,Arboleya:2025ocb} through the skew-whiffing procedure.
The aforementioned issue does not arise for nilmanifolds, as the AdS- and internal curvature scale decouple from the KK scale \cite{Andriot:2016rdd,Andriot:2018tmb}.
In this paper, only for the nilmanifolds with names $\mathfrak{n}_2$, $\mathfrak{n}_4$, $\mathfrak{n}_6$, 37B$_1$, 37C and 37D$_1$, were solutions found where there exists a (partially) homogeneous large flux-limit which leads to large radii, small string coupling and an AdS radius that becomes parametrically larger than any of the radii, and consequentially the lowest eigenvalue of the Laplacian, as the results of refs.~\cite{Andriot:2016rdd,Andriot:2018tmb} indicate.

Therefore, the solutions presented here are, as far as the author knows, the first supersymmetric and parametrically scale-separated solutions of type IIB string theory at large cycle volumes and weak coupling, in the limit where the orientifold planes are smeared.

Interestingly, the mass spectra of some of the solutions, that is, on the solvmanifolds and nilmanifold $\mathfrak{n}_2$, lead to operators in the putative dual field theory with integer conformal dimensions at tree level. The supersymmetric solutions are fully homogeneous under a rescaling of the fluxes. As a consequence, the masses (in AdS units) do not depend on those fluxes. Moreover, all fluxes that are sent to infinity are `trivially unbounded', i.e. they are all independent.
This is very similar to the investigations of refs.~\cite{Conlon:2020wmc,Apers:2022vfp,Ning:2022zqx} for 4d scale-separated vacua. 
However, on all the other backgrounds, the conformal dimensions are not integers. An interesting observation is that for all those vacua, there are scalar fields that have masses in AdS units that are dependent on the fluxes. For all of them except for $\mathfrak{n}_4$, there are also `non-trivially unbounded fluxes', meaning that although they can be sent to infinity, they have to satisfy a constraint. For these cases, there exist flux limits for which some of the scalars become very light compared to the AdS scale. 
For the 3d massive type IIA vacua with scale separation that also features non-trivially unbounded fluxes, the conformal dimensions were also found not to be all integers, as noted in ref.~\cite{Apers:2022zjx}, and it would be interesting to investigate whether there also exist flux limits where some scalars become light. 
Moreover, variants on this model with trivially unbounded fluxes \cite{Farakos:2025bwf}, albeit having massless moduli, also have integer conformal dimensions. Therefore, they seem to support the idea that there is a relation between setups with trivially unbounded fluxes and the presence of integer conformal dimensions and vice versa.
Moreover, it would be interesting to explore further holographic aspects of these solutions in more detail. A starting point could be provided by the recently proposed flux backtracking algorithm \cite{Apers:2025pon}. Especially the solution on the nilmanifold $\mathfrak{n}_2$ seems promising, as this is a compactification to $\mathrm{AdS}_3 \times \mathrm{Nil}_3 \times \mathbb{T}^4$ with $F_3$ and $F_7$ flux and (intersecting) O-planes, which is potentially dual to the world-volume theory of a non-trivial D1-D5 system. Furthermore, this construction also opens up the possibility of going beyond the twisted torus Ansatz by considering compactifications of the type $\mathrm{AdS}_3 \times \mathrm{Nil}_3 \times \mathcal{M}_4$, where $\mathcal{M}_4$ is a four-dimensional Ricci-flat manifold. This is left for future research. 

The 3d solutions presented here share some similarities with their 4d type IIA counterparts, especially the ones in massless type IIA: they rely on intersecting O-planes to stabilise the moduli, and the internal manifold has non-trivial intrinsic torsion. The fact that the O-planes seem to intersect might be a worry, as such setups are not well understood, also not at the world-sheet level. However, ref.~\cite{Junghans:2023yue} showed that for the 4d DGKT vacua \cite{DeWolfe:2005uu,Camara:2005dc} on the specific $\mathbb{T}^6/\mathbb{Z}_3^2$ orbifold, the four sets of O6-planes are not intersecting after blowing up the orbifold singularities. Although it is not entirely clear how the orbifold singularities of the twisted tori should be resolved, one could expect that a blow-up procedure, similar to ordinary $\mathbb{T}^7$'s, might do so. The blow-up moduli could then be stabilised by balancing local RR three-form flux against the bulk seven-form flux, analogous to the blow-up procedure in DGKT~\cite{DeWolfe:2005uu}.
A second reason for worry might be that the O-planes are smeared over the directions transverse to them, not being treated as the local sources that they are. Although computing the local O-plane backreaction for the 3d vacua is a computationally contrived problem here, previous successful attempts for other AdS vacua seem to indicate that there are no fundamental objections to it in principle. Indeed, in massive type IIA, refs.~\cite{Baines:2020dmu,Junghans:2020acz,Marchesano:2020qvg,Emelin:2022cac,Andriot:2023fss,Emelin:2024vug} showed that on various backgrounds without intrinsic torsion, the O-plane backreaction is small and can be computed perturbatively, at first order. Additionally, refs.~\cite{Cribiori:2021djm,VanHemelryck:2024bas} argued the same for backgrounds of massless type IIA with intrinsic torsion, where calculating the backreaction was essential to lift the solutions to M-theory.
Altogether, these results provide evidence that parametrically scale-separated, supersymmetric AdS$_3$ vacua can be constructed in type IIB string theory at large volume and weak coupling in the presence of intersecting orientifolds. 
Although all these solutions relied on specific $\mathbb{Z}_2^3$ orbifolds of seven-dimensional nilmanifolds, it would be interesting to explore other options that yield similar scale-separated solutions in type IIB string theory, like different orbifolds or geometries. Indeed, the ideas presented here suggest that the search for new scale-separated vacua is not over yet.

\section*{Acknowledgements}
I want to thank Adolfo Guarino, Matteo Morittu, Thomas Van Riet, Max Hübner, Magdalena Larfors and especially Álvaro Arboleya and George Tringas for interesting discussions. I am financially supported by the Olle Engkvists Stiftelse.

\appendix
\section{Laplacian spectrum of the ISO(2) solvmanifold}
\label{app:solv_spectrum}
This appendix briefly discusses the eigenfunctions and eigenvalues of the scalar Laplacian of the 3d ISO(2) solvmanifold.\footnote{The full spectrum on the related 3d ISO(1,1) solvmanifold has been studied in ref.~\cite{MR2217282}.} These can easily be generalised to the 7d solvmanifold used in this paper. For this solvmanifold, a twisted torus metric as in eq.~\eqref{eq:metric} can be used with the co-frame one-forms as in eq.~\eqref{eq:one-forms_in_coords}:
\begin{align}
\begin{split}
\label{eq:ISO2_metric}
    \dd s_3^2 =& L_1^2 \left( \cos\left(\omega_a x^7\right) \dd x^1 - \sin\left(\omega_a x^7\right)\dd x^2\right)^2 + L_2^2 \left( \sin\left(\omega_a x^7\right) \dd x^1 + \cos\left(\omega_a x^7\right) \dd x^2\right)^2\\
    &L_7^2 \left(\dd x^7\right)^2\,.
\end{split}
\end{align}
With this metric, one can take the following Ansatz for the eigenfunctions:
\begin{equation}
    Y_{klm} = \e^{2 \pi i (k x^1 + l x^2)} Z_m(x^7)\,.
\end{equation}
Plugging this into the eigenvalue equation
\begin{equation}
    (\Delta - \lambda_{klm}) Y_{klm} = 0\,,
\end{equation}
and constructing the Laplacian with the metric \eqref{eq:ISO2_metric}, one finds that the eigenvalue equation reduces to the Mathieu equation for $Z_m$ \cite{NIST:DLMF}:
\begin{equation}
\label{eq:MathieuEquation}
    Z_{m}'' + \left[a_{klm}-2 q_{kl} \cos\left( 2(\omega_a x^7 + \varphi_{kl})\right)\right] Z_{m} = 0\,,
\end{equation}
where the parameters $a_{klm}$, $q_{kl}$ and $\varphi_{kl}$ are given by:
\begin{gather}
\notag
    a_{klm} = -\frac{L_7^2}{\omega_a^2}\left(\frac{(2\pi)^2(k^2 + l^2)(L_1^2 + L_2^2)}{2L_1^2 L_2^2}+\lambda_{klm}\right), \quad q_{kl}=  -\frac{L_7^2}{\omega_a^2}\left(\frac{(2\pi)^2(k^2 + l^2)(L_1^2 - L_2^2)}{4 L_1^2 L_2^2}\right)\\
    \varphi_{kl} = \frac{1}{2} \arctan \left( \frac{2 kl}{k^2 - l^2} \right)\,.
\end{gather}
The solutions to this equation are periodic in $x^7$ only for specific values of $a_{klm}$, which are called the Mathieu characteristic values, also named $a_m(q)$ and $b_m(q)$, such that 
\begin{equation}
    a_{klm} = a_m(q_{kl}) \quad \text{or} \quad b_m(q_{kl})\,,
\end{equation}
where $m$ is an integer. This discretises the eigenvalue spectrum. The eigenfunctions are then written as (co)sine elliptic functions as follows \cite{NIST:DLMF}:
\begin{equation}
    Z_{m} = \{\mathrm{ce}_m(\omega_a x^7 + \varphi_{kl}, q_{kl}), \mathrm{se}_m(\omega_a x^7 + \varphi_{kl}, q_{kl}) \}\,,
\end{equation}
where $\mathrm{ce}_m$ and $\mathrm{se}_m$ are even and odd functions respectively. 
The eigenvalues of $\mathrm{ce}_m(x,q)$ and $\mathrm{se}_m(x,q)$ are given by $a_m(q)$ and $b_m(q)$ respectively, with the caveat that $\mathrm{se}_0(x,q)$ and $b_0(q)$ do not exist. Additionally, when $m$ is even, then the functions have a periodicity of $\pi$, whereas when they are odd they have antiperiodicity $\pi$. 
However, the case of interest is when $k=l=0$, hence when the eigenfunctions only depend on $x^7$. Then, the eigenvalue equation \eqref{eq:MathieuEquation} reduces to the simple harmonic oscillator and the eigenvalues are:
\begin{equation}
\label{eq:eigenvalue_L7}
    \lambda_{00m} = -\frac{m^2 \omega_a^2}{L_7^2}\,.
\end{equation}
Upon ignoring that $\varphi_{00}$ is ill-defined, this indeed agrees with the Mathieu eigenfunctions when $k=l=0$, as it implies $q_{kl} =0$ (which also happens when $L_1 = L_2$). In that case, $a_m$ and $b_m$ become analytic and simple:
\begin{equation}
    a_m(0) = b_m(0) = m^2,
\end{equation}
and the (co)sine elliptic functions become ordinary (co)sine functions, as expected.
So the tower of eigenvalues given by eq.~\eqref{eq:eigenvalue_L7} only depends on $L_7$. Because $L_7$ is the largest radius, it sets the lowest eigenvalue and the KK scale although $L_7$ decouples from the cycle volumes. Ordinary $\mathbb{Z}_2^n$ orbifolds cannot project out this tower of eigenfunctions. 
Alternatively, when $k\neq 0$ or $l\neq 0$ and $L_2 \gtrsim L_1$, then $q_{kl} \neq 0$, and because $L_7$ is assumed to be much larger than $L_1$ and $L_2$, $q_{kl}$ is large (as $L_2-L_1 = O(1)$). In this limit, both $a_{mkl}$ and $b_{mkl}$ take the same form at leading order for all $m$:
\begin{equation}
    a_{mkl}  = -2 q_{kl} + 2 (2m +1)\;\sqrt{q_{kl}} -\frac{2m(m+1)+1}{4}+ O\left(q_{kl}^{-1/2}\right)\,, 
\end{equation}
This implies that the eigenvalues at leading order are
\begin{align}
\notag
    \lambda_{klm} &= - \frac{\omega_a^2}{L_7^2}a_{klm} -\frac{(2\pi)^2(k^2 + l^2)(L_1^2 + L_2^2)}{2L_1^2 L_2^2}\\
    &\approx 2 \frac{\omega_a^2}{L_7^2}q_{kl}  -\frac{(2\pi)^2(k^2 + l^2)(L_1^2 + L_2^2)}{2L_1^2 L_2^2} = -\frac{(2\pi)^2(k^2 + l^2)}{L_2^2}\,,
\end{align}
which, up to corrections, does not depend on $L_7$, and are bigger in size than $\lambda_{00m}$ as $L_7\gg L_2$.
Finally, it is clear that all these aforementioned eigenfunctions and eigenvalues are also included in the spectrum of the Laplacian of 7d solvmanifold used in this paper. Indeed, the differential equation \eqref{eq:MathieuEquation} would involve a sum of cosines over the 3 different ISO(3) copies, but one can look at modes involving only one copy, as what is done here.
\section{Mass spectra of the nilmanifold compactifications}
\label{app:masses}
In this appendix, the mass spectra of the different nilmanifold compactifications of section~\ref{sec:G2-solutions} are discussed.
\subsubsection*{Nilmanifold $\mathfrak{n}_2$}
This nilmanifold is the simplest of the list, and the scalar masses do not depend on the fluxes as the setup is fully homogeneous. The scalar masses are computed to be
\begin{equation}
    m^2 L_\mathrm{AdS}^2 = \{120, 8,8,8,8,8,8,8\}\,,
\end{equation}
and this leads to integer conformal dimensions, as discussed in the main text.
\subsubsection*{Nilmanifold $\mathfrak{n}_4$}
\begin{figure}[ht]
  \centering
  \begin{minipage}[b]{0.45\textwidth}
    \centering
    \includegraphics[width=\linewidth]{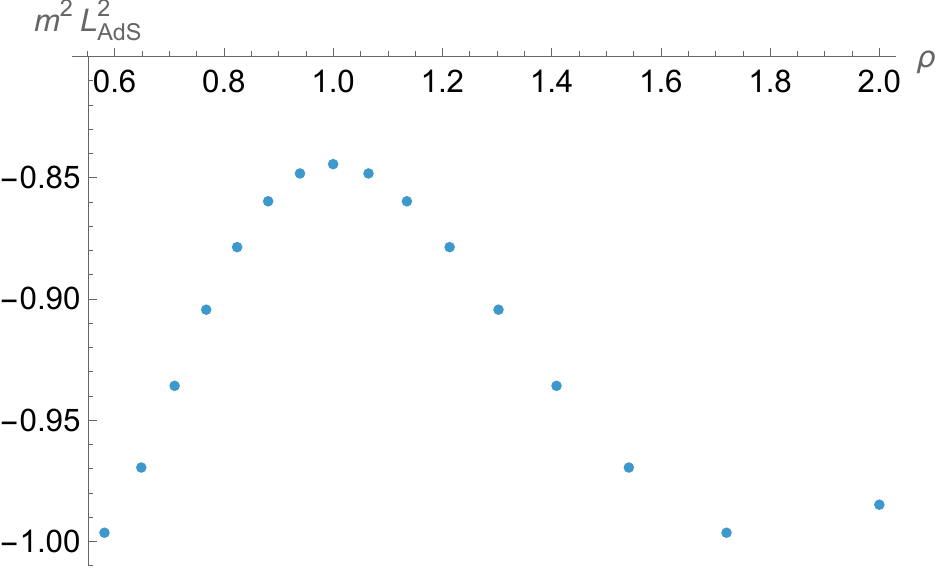}
  \end{minipage}
  \hfill
  \begin{minipage}[b]{0.45\textwidth}
    \centering
    \includegraphics[width=\linewidth]{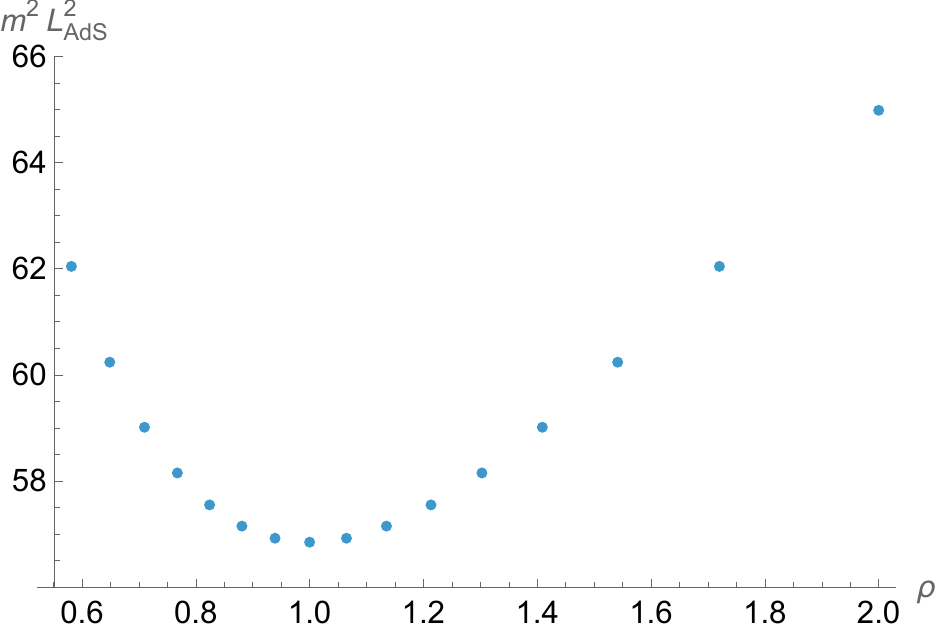}
  \end{minipage}
  \caption{Masses squared for nilmanifold $\mathfrak{n}_4$ for 16 different values of $\rho$, left and right for a scalar with negative and positive mass squared, respectively.}
  \label{fig:masses_n4}
\end{figure}

This nilmanifold is special in the sense that the compactification is only partially homogeneous under flux rescalings, but does not have non-trivially unbounded fluxes. There are 6 out of 8 scalar masses that are flux-independent and are all the same with 
\begin{equation}
    m^2 L_\mathrm{AdS}^2 = 8.
\end{equation}
The other two scalars have mass that depends on $\rho$, but there were only a finite number of consistent values for $\rho$, 16 to be precise. The masses squared are plotted in Figure \ref{fig:masses_n4}.

\subsubsection*{Nilmanifold $\mathfrak{n}_6$}
\begin{figure}[ht]
  \centering
  \begin{minipage}[b]{0.45\textwidth}
    \centering
    \includegraphics[width=\linewidth]{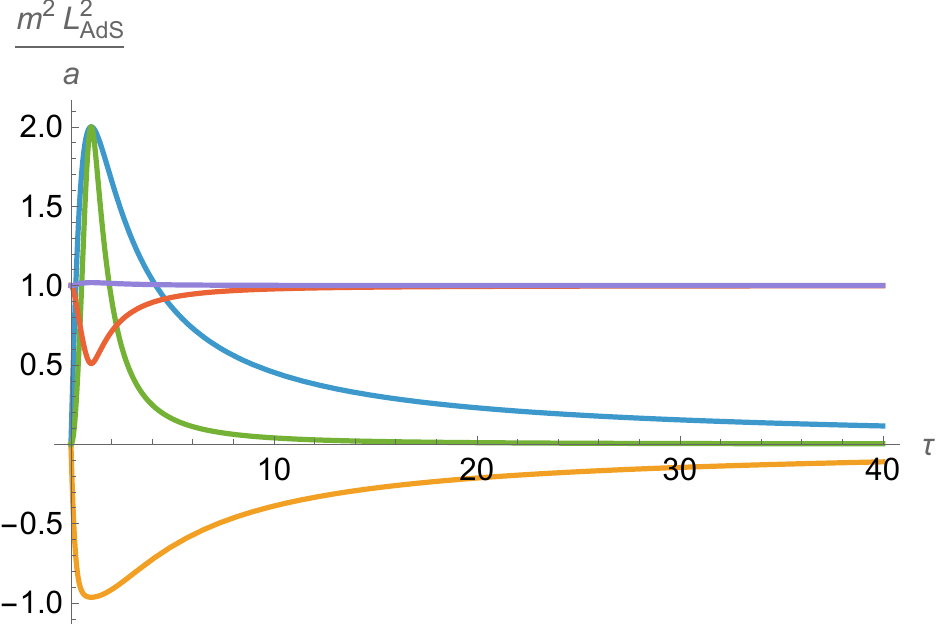}
  \end{minipage}
  \hfill
  \begin{minipage}[b]{0.45\textwidth}
    \centering
    \includegraphics[width=\linewidth]{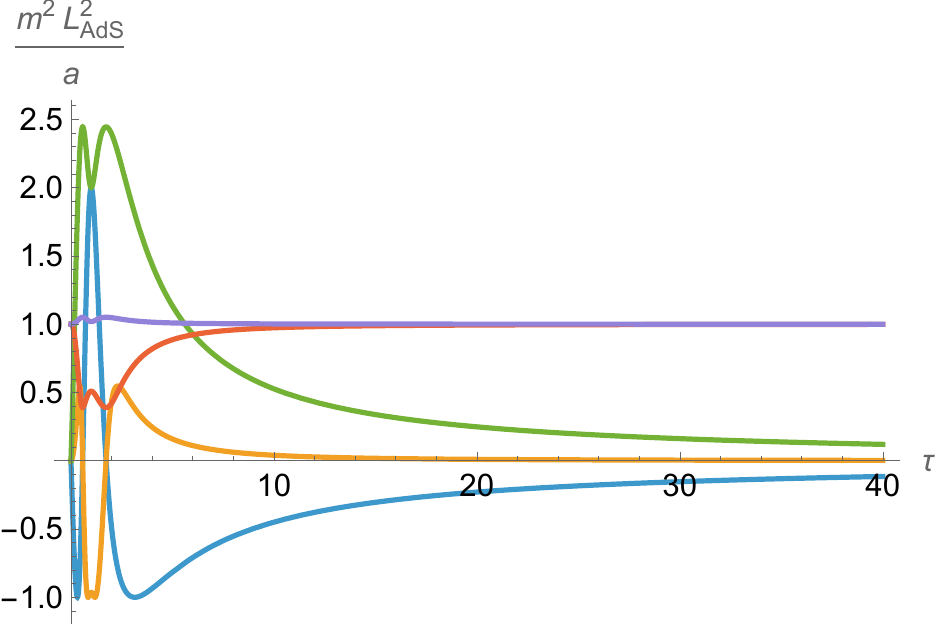}
  \end{minipage}
  \caption{Masses squared for nilmanifold $\mathfrak{n}_6$ dependent on $\tau$, for the $-$ branch (left) and $+$ branch (right). For all lines, $a=1$, except for the red and purple, which are normalised with $a=8$ and $a=48$ respectively.}
  \label{fig:masses_n6}
\end{figure}

For this nilmanifold, there are non-trivially unbounded fluxes, and the solutions are parametrised here by a parameter $\tau$ introduced first in eq.~\eqref{eq:moduli_Ansatz_n6}.
Irrespective of $\tau$, the trace of the mass matrix is 80, meaning that the total sum of the masses should be 80. 
For this nilmanifold, 3 out of 8 scalar masses are all the same with 
\begin{equation}
    m^2 L_\mathrm{AdS}^2 = 8.
\end{equation}
The other 6 scalars have masses that depend on $\rho$ and $\tau$, which are plotted in Figure \ref{fig:masses_n6} corresponding to the two branches of $\rho$ (see eq.~\eqref{eq:n6_rhotau_sol}). 
In the large $\tau$ limit, three of the masses asymptote to 0, whereas the other two go to the values 8 and 48.
\begin{equation}
m^2 L_\mathrm{AdS}^2 \underset{\tau\to \infty}{=}\left\{48,8,8,8,8,0,0,0\right\} + O(\tau^{-1})\,.
\end{equation}
Even at large $\tau$, one can obtain scale separation in the well-controlled regime, as long as the flux condition~\eqref{eq:flux_limit_n6} is replaced by
\begin{equation}
    f_3^2 \tau^2 \ll f_7 \ll f_3^4, \qquad \tau^2 \ll f_3\,.
\end{equation}

\subsubsection*{Nilmanifold 37B$_1$}
\begin{figure}[ht]
  \centering
  \begin{minipage}[b]{0.45\textwidth}
    \centering
    \includegraphics[width=\linewidth]{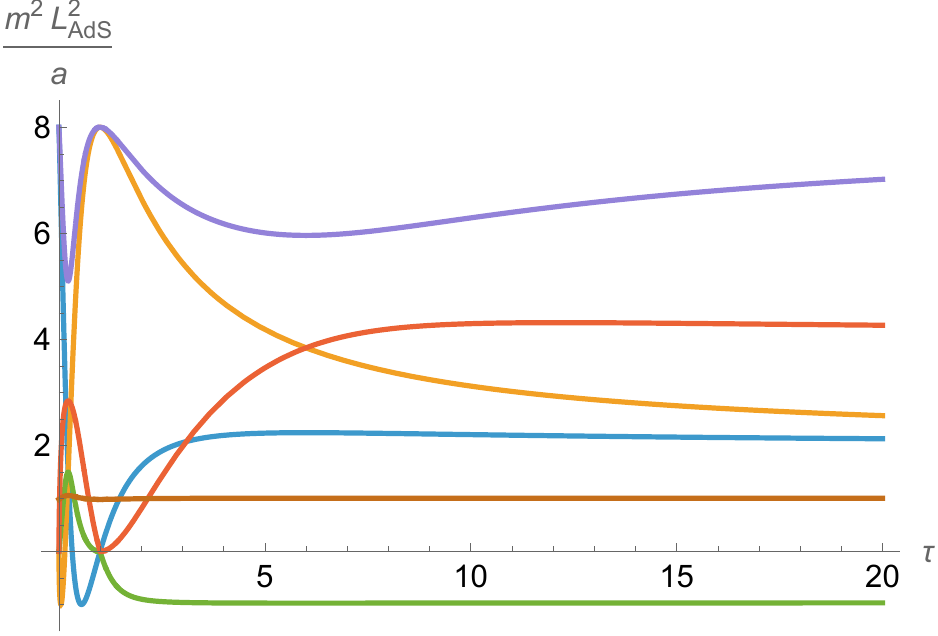}
  \end{minipage}
  \hfill
  \begin{minipage}[b]{0.45\textwidth}
    \centering
    \includegraphics[width=\linewidth]{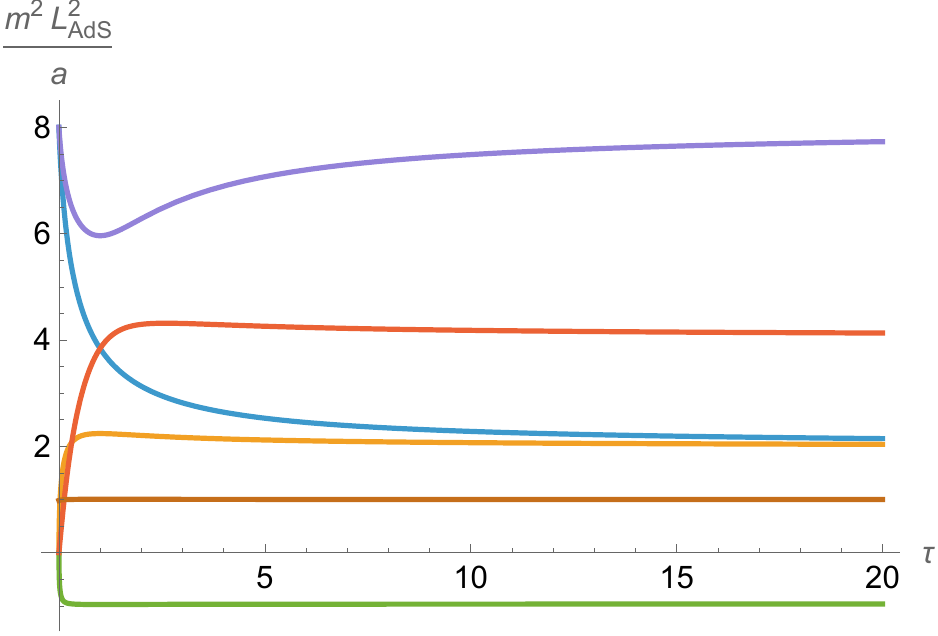}
  \end{minipage}
  \caption{Masses squared for nilmanifold 37B$_1$ dependent on $\tau$, for the $-$ branch (left) and $+$ branch (right). For all lines, $a=1$, except for the brown one, which is normalised with $a=4$. The plot of the $-$ branch starts from $\tau =1$, as the solution is only consistent for this branch when $\tau>1$.}
  \label{fig:masses_37B1}
\end{figure}

For nilmanifold 37B$_1$, the trace of the mass matrix is also equal to 80. 
Two out of eight scalar masses are the same and independent of $\tau$ (as introduced in eq.~\eqref{eq:moduli_Ansatz_37B1}: 
\begin{equation}
    m^2 L_\mathrm{AdS}^2 = 8.
\end{equation}
The other 5 scalars have mass that depends on $\rho$, which are plotted in Figure \ref{fig:masses_37B1}, for both the $+$ and $-$ branches of eq.~\eqref{eq:37B1_rhotau_sol}.
In the large $\tau$-limit, three scalars become light, and the asymptotic values of all scalars are:
\begin{equation}
m^2 L_\mathrm{AdS}^2 \underset{\tau\to \infty}{=}\left\{48,8,8,8,8,0,0,0\right\} + O(\tau^{-1})\,.
\end{equation}
A well-controlled and scale-separated regime still exists in this limit, as long as the conditions on the fluxes \eqref{eq:flux_limit_37B1} are modified to
\begin{equation}
    f_7 \tau^2\ll F_{235}^4 \ll f_7^2 \tau^{-4}, \tau^2 \ll f_7\,.
\end{equation}

\subsubsection*{Nilmanifold 37C}
\begin{figure}[ht]
  \centering
  \begin{minipage}[b]{0.45\textwidth}
    \centering
    \includegraphics[width=\linewidth]{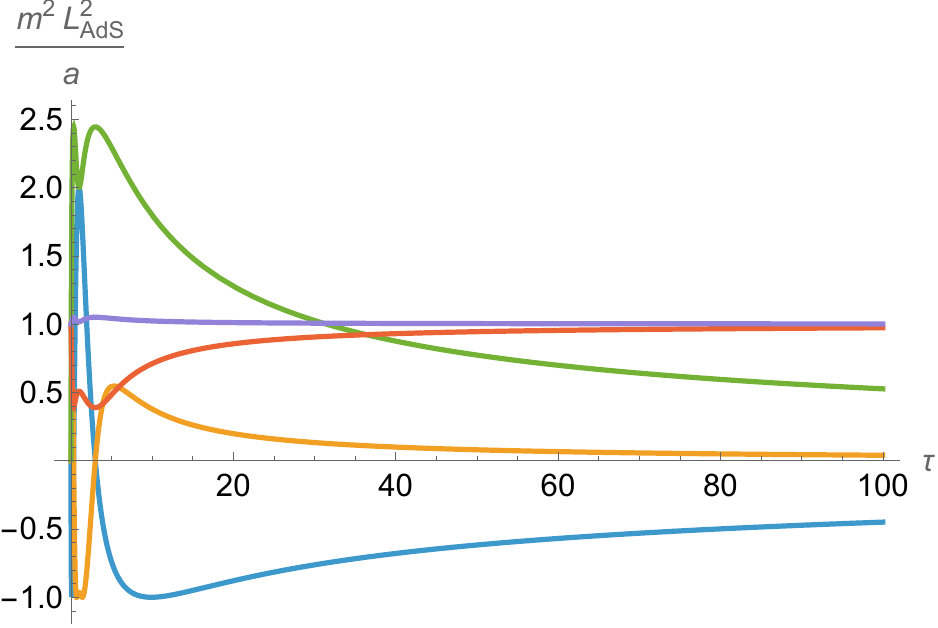}
  \end{minipage}
  \hfill
  \begin{minipage}[b]{0.45\textwidth}
    \centering
    \includegraphics[width=\linewidth]{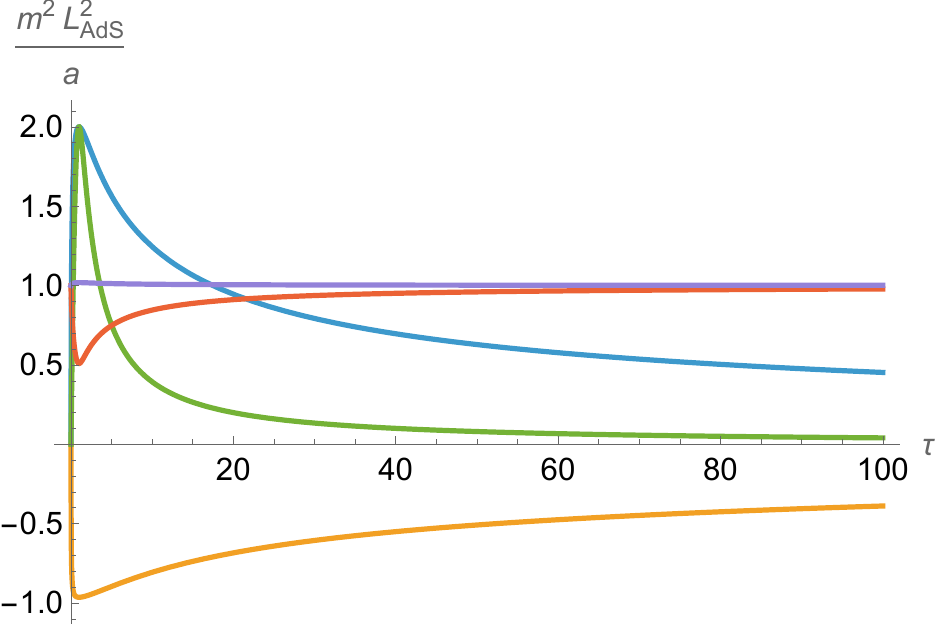}
  \end{minipage}
  \caption{Masses squared for nilmanifold 37C dependent on $\tau$, for the $-$ branch (left) and $+$ branch (right). For all lines, $a=1$, except for the red and purple, which are normalised with $a=8$ and $a=48$ respectively.}
  \label{fig:masses_37C}
\end{figure}

Many qualitative features of the mass spectra for the previous nilmanifolds persist for nilmanifold 37C: the trace of the mass matrix is equal to 80, and there are also 3 out of 8 scalar masses that are all the same and independent of $\tau$ and $\rho$ as introduced in eq.~\eqref{eq:moduli_Ansatz_37C}: 
\begin{equation}
    m^2 L_\mathrm{AdS}^2 = 8.
\end{equation}
The other 5 scalars have masses that depend on $\rho$ and $\tau$, and are plotted in Figure \ref{fig:masses_37C} for the two branches of \eqref{eq:37C_rhotau_sol}. 
In the large $\tau$ limit, three of the masses asymptote to 0, whereas the other two go to the values 8 and 48.
\begin{equation}
m^2 L_\mathrm{AdS}^2 \underset{\tau\to \infty}{=}\left\{48,8,8,8,8,0,0,0\right\} + O(\tau^{-1})\,.
\end{equation}
Also in this limit, one can obtain scale-separated and well-controlled solutions, as long as the flux limit \eqref{eq:flux_limit_37C} is modified to
\begin{equation}
    F_{235}^2 \tau^4 \ll f_7 \ll F_{235}^4 \tau^2\,, \qquad  \tau^6 \ll f_7\,.
\end{equation}

\subsubsection*{Nilmanifold 37D$_1$}
\begin{figure}[ht]
  \centering
  \begin{minipage}[b]{0.45\textwidth}
    \centering
    \includegraphics[width=\linewidth]{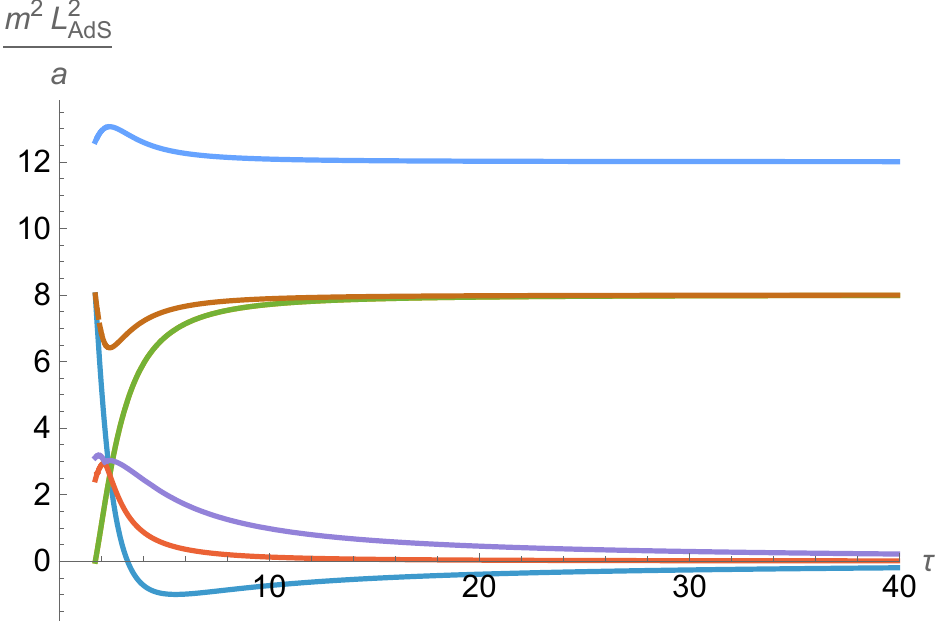}
  \end{minipage}
  \hfill
  \begin{minipage}[b]{0.45\textwidth}
    \centering
    \includegraphics[width=\linewidth]{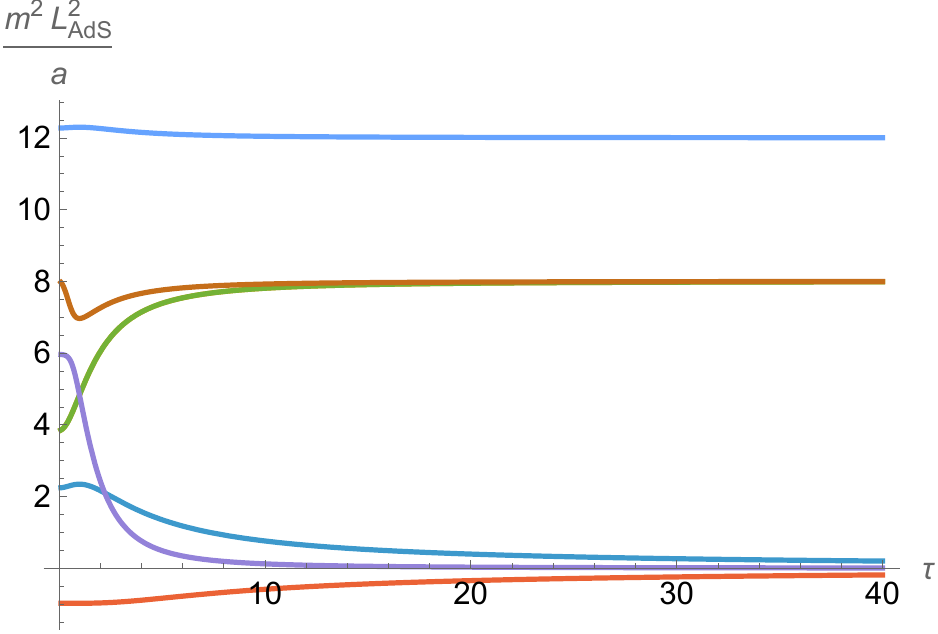}
  \end{minipage}
  \caption{Masses squared dependent on $\tau$, for the $-$ branch (left) and $+$ branch (right). For all lines, $a=1$, except for the upper curve in light blue, which is normalised with $a=4$. The green curve (third highest up) has multiplicity 2. For the $-$ branch, $\rho$ is only positive if $\tau > \sqrt{3}$ and hence the plot starts from this value.}
  \label{fig:masses_37D1}
\end{figure}

Also, this last nilmanifold shares some features of the mass spectra with the previous nilmanifolds: the trace of the mass matrix is equal to 80, but this time there is only one scalar mass that is independent of $\tau$ and $\rho$ as introduced in eq.~\eqref{eq:moduli_Ansatz_37D1} with
\begin{equation}
    m^2 L_\mathrm{AdS}^2 = 8.
\end{equation}
The other 7 scalars have mass that depends on $\tau$, which are plotted in Figure \ref{fig:masses_37D1} corresponding to the two branches of \eqref{eq:37D1_rhotau_sol}.
In the large $\tau$ limit, three of the masses asymptote to 0, whereas the other four go to the values 8 and 48.
\begin{equation}
m^2 L_\mathrm{AdS}^2 \underset{\tau\to \infty}{=}\left\{48,8,8,8,8,0,0,0\right\} + O(\tau^{-1})\,.
\end{equation}
Also here, one can obtain scale separation in the well-controlled regime for large $\tau$, as long as the flux condition~\eqref{eq:flux_limit_37D1} is replaced by
\begin{equation}
    F_{235}^2 \tau^6 \ll f_7 \ll F_{235}^4 \tau^2\,, \qquad \tau^2 \ll F_{235}\,.
\end{equation}

\bibliographystyle{JHEP}
\bibliography{AllRefs.bib}

\providecommand{\href}[2]{#2}\begingroup\raggedright\begin{thebibliography}{10}

\bibitem{Cribiori:2022trc}
N.~Cribiori and G.~Dall'Agata, \emph{{Weak gravity versus scale separation}},
  \href{https://doi.org/10.1007/JHEP06(2022)006}{\emph{JHEP} {\bfseries 06}
  (2022) 006} [\href{https://arxiv.org/abs/2203.05559}{{\ttfamily
  2203.05559}}].

\bibitem{Bobev:2023dwx}
N.~Bobev, M.~David, J.~Hong, V.~Reys and X.~Zhang, \emph{{A compendium of
  logarithmic corrections in AdS/CFT}},
  \href{https://doi.org/10.1007/JHEP04(2024)020}{\emph{JHEP} {\bfseries 04}
  (2024) 020} [\href{https://arxiv.org/abs/2312.08909}{{\ttfamily
  2312.08909}}].

\bibitem{Cribiori:2023ihv}
N.~Cribiori and C.~Montella, \emph{{Quantum gravity constraints on scale
  separation and de Sitter in five dimensions}},
  \href{https://doi.org/10.1007/JHEP05(2023)178}{\emph{JHEP} {\bfseries 05}
  (2023) 178} [\href{https://arxiv.org/abs/2303.04162}{{\ttfamily
  2303.04162}}].

\bibitem{Perlmutter:2024noo}
E.~Perlmutter, \emph{{Rigorous Holographic Bound on AdS Scale Separation}},
  \href{https://doi.org/10.1103/PhysRevLett.133.061601}{\emph{Phys. Rev. Lett.}
  {\bfseries 133} (2024) 061601}
  [\href{https://arxiv.org/abs/2402.19358}{{\ttfamily 2402.19358}}].

\bibitem{Gautason:2015tig}
F.~F. Gautason, M.~Schillo, T.~Van~Riet and M.~Williams, \emph{{Remarks on
  scale separation in flux vacua}},
  \href{https://doi.org/10.1007/JHEP03(2016)061}{\emph{JHEP} {\bfseries 03}
  (2016) 061} [\href{https://arxiv.org/abs/1512.00457}{{\ttfamily
  1512.00457}}].

\bibitem{Collins:2022nux}
T.~C. Collins, D.~Jafferis, C.~Vafa, K.~Xu and S.-T. Yau, \emph{{On Upper
  Bounds in Dimension Gaps of CFT's}},
  \href{https://arxiv.org/abs/2201.03660}{{\ttfamily 2201.03660}}.

\bibitem{Lust:2019zwm}
D.~L{\"u}st, E.~Palti and C.~Vafa, \emph{{AdS and the Swampland}},
  \href{https://doi.org/10.1016/j.physletb.2019.134867}{\emph{Phys. Lett. B}
  {\bfseries 797} (2019) 134867}
  [\href{https://arxiv.org/abs/1906.05225}{{\ttfamily 1906.05225}}].

\bibitem{Coudarchet:2023mfs}
T.~Coudarchet, \emph{{Hiding the extra dimensions: A review on scale separation
  in string theory}},
  \href{https://doi.org/10.1016/j.physrep.2024.02.003}{\emph{Phys. Rept.}
  {\bfseries 1064} (2024) 1}
  [\href{https://arxiv.org/abs/2311.12105}{{\ttfamily 2311.12105}}].

\bibitem{DeWolfe:2005uu}
O.~DeWolfe, A.~Giryavets, S.~Kachru and W.~Taylor, \emph{{Type IIA moduli
  stabilization}},
  \href{https://doi.org/10.1088/1126-6708/2005/07/066}{\emph{JHEP} {\bfseries
  07} (2005) 066} [\href{https://arxiv.org/abs/hep-th/0505160}{{\ttfamily
  hep-th/0505160}}].

\bibitem{Camara:2005dc}
P.~G. Camara, A.~Font and L.~E. Ibanez, \emph{{Fluxes, moduli fixing and
  MSSM-like vacua in a simple IIA orientifold}},
  \href{https://doi.org/10.1088/1126-6708/2005/09/013}{\emph{JHEP} {\bfseries
  09} (2005) 013} [\href{https://arxiv.org/abs/hep-th/0506066}{{\ttfamily
  hep-th/0506066}}].

\bibitem{Cribiori:2021djm}
N.~Cribiori, D.~Junghans, V.~Van~Hemelryck, T.~Van~Riet and T.~Wrase,
  \emph{{Scale-separated AdS4 vacua of IIA orientifolds and M-theory}},
  \href{https://doi.org/10.1103/PhysRevD.104.126014}{\emph{Phys. Rev. D}
  {\bfseries 104} (2021) 126014}
  [\href{https://arxiv.org/abs/2107.00019}{{\ttfamily 2107.00019}}].

\bibitem{Carrasco:2023hta}
R.~Carrasco, T.~Coudarchet, F.~Marchesano and D.~Prieto, \emph{{New families of
  scale separated vacua}},
  \href{https://doi.org/10.1007/JHEP11(2023)094}{\emph{JHEP} {\bfseries 11}
  (2023) 094} [\href{https://arxiv.org/abs/2309.00043}{{\ttfamily
  2309.00043}}].

\bibitem{VanHemelryck:2024bas}
V.~Van~Hemelryck, \emph{{Weak G2 manifolds and scale separation in M-theory
  from type IIA backgrounds}},
  \href{https://doi.org/10.1103/PhysRevD.110.106013}{\emph{Phys. Rev. D}
  {\bfseries 110} (2024) 106013}
  [\href{https://arxiv.org/abs/2408.16609}{{\ttfamily 2408.16609}}].

\bibitem{Farakos:2020phe}
F.~Farakos, G.~Tringas and T.~Van~Riet, \emph{{No-scale and scale-separated
  flux vacua from IIA on G2 orientifolds}},
  \href{https://doi.org/10.1140/epjc/s10052-020-8247-5}{\emph{Eur. Phys. J. C}
  {\bfseries 80} (2020) 659}
  [\href{https://arxiv.org/abs/2005.05246}{{\ttfamily 2005.05246}}].

\bibitem{VanHemelryck:2022ynr}
V.~Van~Hemelryck, \emph{{Scale-Separated AdS3 Vacua from G2-Orientifolds Using
  Bispinors}}, \href{https://doi.org/10.1002/prop.202200128}{\emph{Fortsch.
  Phys.} {\bfseries 70} (2022) 2200128}
  [\href{https://arxiv.org/abs/2207.14311}{{\ttfamily 2207.14311}}].

\bibitem{Farakos:2023nms}
F.~Farakos, M.~Morittu and G.~Tringas, \emph{{On/off scale separation}},
  \href{https://doi.org/10.1007/JHEP10(2023)067}{\emph{JHEP} {\bfseries 10}
  (2023) 067} [\href{https://arxiv.org/abs/2304.14372}{{\ttfamily
  2304.14372}}].

\bibitem{Farakos:2023wps}
F.~Farakos and M.~Morittu, \emph{{Scale-separated AdS$_3\times $S$^1$ vacua
  from IIA orientifolds}},
  \href{https://doi.org/10.1140/epjc/s10052-024-12427-z}{\emph{Eur. Phys. J. C}
  {\bfseries 84} (2024) 98} [\href{https://arxiv.org/abs/2311.08991}{{\ttfamily
  2311.08991}}].

\bibitem{Tringas:2023vzn}
G.~Tringas, \emph{{Anisotropic scale-separated AdS$_{4}$ flux vacua}},
  \href{https://doi.org/10.1007/JHEP04(2025)151}{\emph{JHEP} {\bfseries 04}
  (2025) 151} [\href{https://arxiv.org/abs/2309.16542}{{\ttfamily
  2309.16542}}].

\bibitem{Andriot:2025cyi}
D.~Andriot, N.~Cribiori and T.~Van~Riet, \emph{{Scale separation, rolling
  solutions, and entropy bounds}},
  \href{https://doi.org/10.1103/5rkw-5qfk}{\emph{Phys. Rev. D} {\bfseries 112}
  (2025) 026028} [\href{https://arxiv.org/abs/2504.08634}{{\ttfamily
  2504.08634}}].

\bibitem{Conlon:2021cjk}
J.~P. Conlon, S.~Ning and F.~Revello, \emph{{Exploring the holographic
  Swampland}}, \href{https://doi.org/10.1007/JHEP04(2022)117}{\emph{JHEP}
  {\bfseries 04} (2022) 117}
  [\href{https://arxiv.org/abs/2110.06245}{{\ttfamily 2110.06245}}].

\bibitem{Apers:2022tfm}
F.~Apers, J.~P. Conlon, S.~Ning and F.~Revello, \emph{{Integer conformal
  dimensions for type IIa flux vacua}},
  \href{https://doi.org/10.1103/PhysRevD.105.106029}{\emph{Phys. Rev. D}
  {\bfseries 105} (2022) 106029}
  [\href{https://arxiv.org/abs/2202.09330}{{\ttfamily 2202.09330}}].

\bibitem{Quirant:2022fpn}
J.~Quirant, \emph{{Noninteger conformal dimensions for type IIA flux vacua}},
  \href{https://doi.org/10.1103/PhysRevD.106.066017}{\emph{Phys. Rev. D}
  {\bfseries 106} (2022) 066017}
  [\href{https://arxiv.org/abs/2204.00014}{{\ttfamily 2204.00014}}].

\bibitem{Plauschinn:2022ztd}
E.~Plauschinn, \emph{{Mass spectrum of type IIB flux compactifications
  {\textemdash} comments on AdS vacua and conformal dimensions}},
  \href{https://doi.org/10.1007/JHEP02(2023)257}{\emph{JHEP} {\bfseries 02}
  (2023) 257} [\href{https://arxiv.org/abs/2210.04528}{{\ttfamily
  2210.04528}}].

\bibitem{Apers:2022zjx}
F.~Apers, M.~Montero, T.~Van~Riet and T.~Wrase, \emph{{Comments on classical
  AdS flux vacua with scale separation}},
  \href{https://doi.org/10.1007/JHEP05(2022)167}{\emph{JHEP} {\bfseries 05}
  (2022) 167} [\href{https://arxiv.org/abs/2202.00682}{{\ttfamily
  2202.00682}}].

\bibitem{Baines:2020dmu}
S.~Baines and T.~Van~Riet, \emph{{Smearing orientifolds in flux
  compactifications can be OK}},
  \href{https://doi.org/10.1088/1361-6382/aba8e0}{\emph{Class. Quant. Grav.}
  {\bfseries 37} (2020) 195015}
  [\href{https://arxiv.org/abs/2005.09501}{{\ttfamily 2005.09501}}].

\bibitem{Junghans:2020acz}
D.~Junghans, \emph{{O-Plane Backreaction and Scale Separation in Type IIA Flux
  Vacua}}, \href{https://doi.org/10.1002/prop.202000040}{\emph{Fortsch. Phys.}
  {\bfseries 68} (2020) 2000040}
  [\href{https://arxiv.org/abs/2003.06274}{{\ttfamily 2003.06274}}].

\bibitem{Marchesano:2020qvg}
F.~Marchesano, E.~Palti, J.~Quirant and A.~Tomasiello, \emph{{On supersymmetric
  AdS$_{4}$ orientifold vacua}},
  \href{https://doi.org/10.1007/JHEP08(2020)087}{\emph{JHEP} {\bfseries 08}
  (2020) 087} [\href{https://arxiv.org/abs/2003.13578}{{\ttfamily
  2003.13578}}].

\bibitem{Andriot:2023fss}
D.~Andriot and G.~Tringas, \emph{{Extensions of a scale-separated AdS$_{4}$
  solution and their mass spectrum}},
  \href{https://doi.org/10.1007/JHEP01(2024)008}{\emph{JHEP} {\bfseries 01}
  (2024) 008} [\href{https://arxiv.org/abs/2310.06115}{{\ttfamily
  2310.06115}}].

\bibitem{Emelin:2024vug}
M.~Emelin, \emph{{Consistency conditions for O-plane unsmearing from
  second-order perturbation theory}},
  \href{https://doi.org/10.1007/JHEP12(2024)025}{\emph{JHEP} {\bfseries 12}
  (2024) 025} [\href{https://arxiv.org/abs/2407.12717}{{\ttfamily
  2407.12717}}].

\bibitem{Junghans:2023yue}
D.~Junghans, \emph{{A note on O6 intersections in AdS flux vacua}},
  \href{https://doi.org/10.1007/JHEP02(2024)126}{\emph{JHEP} {\bfseries 02}
  (2024) 126} [\href{https://arxiv.org/abs/2310.17695}{{\ttfamily
  2310.17695}}].

\bibitem{Montero:2024qtz}
M.~Montero and I.~Valenzuela, \emph{{Quantum corrections to DGKT and the Weak
  Gravity Conjecture}},
  \href{https://doi.org/10.1007/JHEP07(2025)057}{\emph{JHEP} {\bfseries 07}
  (2025) 057} [\href{https://arxiv.org/abs/2412.00189}{{\ttfamily
  2412.00189}}].

\bibitem{Arkani-Hamed:2006emk}
N.~Arkani-Hamed, L.~Motl, A.~Nicolis and C.~Vafa, \emph{{The String landscape,
  black holes and gravity as the weakest force}},
  \href{https://doi.org/10.1088/1126-6708/2007/06/060}{\emph{JHEP} {\bfseries
  06} (2007) 060} [\href{https://arxiv.org/abs/hep-th/0601001}{{\ttfamily
  hep-th/0601001}}].

\bibitem{Kachru:2003aw}
S.~Kachru, R.~Kallosh, A.~D. Linde and S.~P. Trivedi, \emph{{De Sitter vacua in
  string theory}},
  \href{https://doi.org/10.1103/PhysRevD.68.046005}{\emph{Phys. Rev. D}
  {\bfseries 68} (2003) 046005}
  [\href{https://arxiv.org/abs/hep-th/0301240}{{\ttfamily hep-th/0301240}}].

\bibitem{Conlon:2005ki}
J.~P. Conlon, F.~Quevedo and K.~Suruliz, \emph{{Large-volume flux
  compactifications: Moduli spectrum and D3/D7 soft supersymmetry breaking}},
  \href{https://doi.org/10.1088/1126-6708/2005/08/007}{\emph{JHEP} {\bfseries
  08} (2005) 007} [\href{https://arxiv.org/abs/hep-th/0505076}{{\ttfamily
  hep-th/0505076}}].

\bibitem{Balasubramanian:2005zx}
V.~Balasubramanian, P.~Berglund, J.~P. Conlon and F.~Quevedo,
  \emph{{Systematics of moduli stabilisation in Calabi-Yau flux
  compactifications}},
  \href{https://doi.org/10.1088/1126-6708/2005/03/007}{\emph{JHEP} {\bfseries
  03} (2005) 007} [\href{https://arxiv.org/abs/hep-th/0502058}{{\ttfamily
  hep-th/0502058}}].

\bibitem{Demirtas:2019sip}
M.~Demirtas, M.~Kim, L.~Mcallister and J.~Moritz, \emph{{Vacua with Small Flux
  Superpotential}},
  \href{https://doi.org/10.1103/PhysRevLett.124.211603}{\emph{Phys. Rev. Lett.}
  {\bfseries 124} (2020) 211603}
  [\href{https://arxiv.org/abs/1912.10047}{{\ttfamily 1912.10047}}].

\bibitem{Demirtas:2020ffz}
M.~Demirtas, M.~Kim, L.~McAllister and J.~Moritz, \emph{{Conifold Vacua with
  Small Flux Superpotential}},
  \href{https://doi.org/10.1002/prop.202000085}{\emph{Fortsch. Phys.}
  {\bfseries 68} (2020) 2000085}
  [\href{https://arxiv.org/abs/2009.03312}{{\ttfamily 2009.03312}}].

\bibitem{Demirtas:2021nlu}
M.~Demirtas, M.~Kim, L.~McAllister, J.~Moritz and A.~Rios-Tascon, \emph{{Small
  cosmological constants in string theory}},
  \href{https://doi.org/10.1007/JHEP12(2021)136}{\emph{JHEP} {\bfseries 12}
  (2021) 136} [\href{https://arxiv.org/abs/2107.09064}{{\ttfamily
  2107.09064}}].

\bibitem{Demirtas:2021ote}
M.~Demirtas, M.~Kim, L.~McAllister, J.~Moritz and A.~Rios-Tascon,
  \emph{{Exponentially Small Cosmological Constant in String Theory}},
  \href{https://doi.org/10.1103/PhysRevLett.128.011602}{\emph{Phys. Rev. Lett.}
  {\bfseries 128} (2022) 011602}
  [\href{https://arxiv.org/abs/2107.09065}{{\ttfamily 2107.09065}}].

\bibitem{McAllister:2024lnt}
L.~McAllister, J.~Moritz, R.~Nally and A.~Schachner, \emph{{Candidate de Sitter
  vacua}}, \href{https://doi.org/10.1103/PhysRevD.111.086015}{\emph{Phys. Rev.
  D} {\bfseries 111} (2025) 086015}
  [\href{https://arxiv.org/abs/2406.13751}{{\ttfamily 2406.13751}}].

\bibitem{Bena:2018fqc}
I.~Bena, E.~Dudas, M.~Gra{\~n}a and S.~L{\"u}st, \emph{{Uplifting Runaways}},
  \href{https://doi.org/10.1002/prop.201800100}{\emph{Fortsch. Phys.}
  {\bfseries 67} (2019) 1800100}
  [\href{https://arxiv.org/abs/1809.06861}{{\ttfamily 1809.06861}}].

\bibitem{Gao:2020xqh}
X.~Gao, A.~Hebecker and D.~Junghans, \emph{{Control issues of KKLT}},
  \href{https://doi.org/10.1002/prop.202000089}{\emph{Fortsch. Phys.}
  {\bfseries 68} (2020) 2000089}
  [\href{https://arxiv.org/abs/2009.03914}{{\ttfamily 2009.03914}}].

\bibitem{Bena:2020xrh}
I.~Bena, J.~Bl\r{a}b\"ack, M.~Gra\~na and S.~L\"ust, \emph{{The tadpole
  problem}}, \href{https://doi.org/10.1007/JHEP11(2021)223}{\emph{JHEP}
  {\bfseries 11} (2021) 223}
  [\href{https://arxiv.org/abs/2010.10519}{{\ttfamily 2010.10519}}].

\bibitem{Gao:2022fdi}
X.~Gao, A.~Hebecker, S.~Schreyer and V.~Venken, \emph{{The LVS parametric
  tadpole constraint}},
  \href{https://doi.org/10.1007/JHEP07(2022)056}{\emph{JHEP} {\bfseries 07}
  (2022) 056} [\href{https://arxiv.org/abs/2202.04087}{{\ttfamily
  2202.04087}}].

\bibitem{Gao:2022uop}
X.~Gao, A.~Hebecker, S.~Schreyer and V.~Venken, \emph{{Loops, local corrections
  and warping in the LVS and other type IIB models}},
  \href{https://doi.org/10.1007/JHEP09(2022)091}{\emph{JHEP} {\bfseries 09}
  (2022) 091} [\href{https://arxiv.org/abs/2204.06009}{{\ttfamily
  2204.06009}}].

\bibitem{Lust:2022lfc}
S.~L\"ust, C.~Vafa, M.~Wiesner and K.~Xu, \emph{{Holography and the KKLT
  scenario}}, \href{https://doi.org/10.1007/JHEP10(2022)188}{\emph{JHEP}
  {\bfseries 10} (2022) 188}
  [\href{https://arxiv.org/abs/2204.07171}{{\ttfamily 2204.07171}}].

\bibitem{Caviezel:2009tu}
C.~Caviezel, T.~Wrase and M.~Zagermann, \emph{{Moduli Stabilization and
  Cosmology of Type IIB on SU(2)-Structure Orientifolds}},
  \href{https://doi.org/10.1007/JHEP04(2010)011}{\emph{JHEP} {\bfseries 04}
  (2010) 011} [\href{https://arxiv.org/abs/0912.3287}{{\ttfamily 0912.3287}}].

\bibitem{Petrini:2013ika}
M.~Petrini, G.~Solard and T.~Van~Riet, \emph{{AdS vacua with scale separation
  from IIB supergravity}},
  \href{https://doi.org/10.1007/JHEP11(2013)010}{\emph{JHEP} {\bfseries 11}
  (2013) 010} [\href{https://arxiv.org/abs/1308.1265}{{\ttfamily 1308.1265}}].

\bibitem{Tringas:2025uyg}
G.~Tringas and T.~Wrase, \emph{{Scale separation from O-planes}},
  \href{https://arxiv.org/abs/2504.15436}{{\ttfamily 2504.15436}}.

\bibitem{Passias:2020ubv}
A.~Passias and D.~Prins, \emph{{On supersymmetric AdS$_{3}$ solutions of Type
  II}}, \href{https://doi.org/10.1007/JHEP08(2021)168}{\emph{JHEP} {\bfseries
  08} (2021) 168} [\href{https://arxiv.org/abs/2011.00008}{{\ttfamily
  2011.00008}}].

\bibitem{Emelin:2021gzx}
M.~Emelin, F.~Farakos and G.~Tringas, \emph{{Three-dimensional flux vacua from
  IIB on co-calibrated G2 orientifolds}},
  \href{https://doi.org/10.1140/epjc/s10052-021-09261-y}{\emph{Eur. Phys. J. C}
  {\bfseries 81} (2021) 456}
  [\href{https://arxiv.org/abs/2103.03282}{{\ttfamily 2103.03282}}].

\bibitem{DallAgata:2005zlf}
G.~Dall'Agata and N.~Prezas, \emph{{Scherk-Schwarz reduction of M-theory on
  G2-manifolds with fluxes}},
  \href{https://doi.org/10.1088/1126-6708/2005/10/103}{\emph{JHEP} {\bfseries
  10} (2005) 103} [\href{https://arxiv.org/abs/hep-th/0509052}{{\ttfamily
  hep-th/0509052}}].

\bibitem{Andriot:2015sia}
D.~Andriot, \emph{{New supersymmetric vacua on solvmanifolds}},
  \href{https://doi.org/10.1007/JHEP02(2016)112}{\emph{JHEP} {\bfseries 02}
  (2016) 112} [\href{https://arxiv.org/abs/1507.00014}{{\ttfamily
  1507.00014}}].

\bibitem{Manero2020315}
V.~Manero, \emph{Compact solvmanifolds with calibrated and cocalibrated g 2
  -structures},
  \href{https://doi.org/10.1007/s00229-019-01133-w}{\emph{Manuscripta
  Mathematica} {\bfseries 162} (2020) 315 }.

\bibitem{Arboleya:2024vnp}
A.~Arboleya, A.~Guarino and M.~Morittu, \emph{{Type II orientifold flux vacua
  in 3D}}, \href{https://doi.org/10.1007/JHEP12(2024)087}{\emph{JHEP}
  {\bfseries 12} (2024) 087}
  [\href{https://arxiv.org/abs/2408.01403}{{\ttfamily 2408.01403}}].

\bibitem{Arboleya:2025ocb}
A.~Arboleya, A.~Guarino and M.~Morittu, \emph{{On type IIB AdS$_{3}$ flux vacua
  with scale separation and integer conformal dimensions}},  in \emph{{24th
  Hellenic School and Workshops on Elementary Particle Physics and Gravity}},
  4, 2025, \href{https://arxiv.org/abs/2504.21508}{{\ttfamily 2504.21508}}.

\bibitem{MR2811660}
D.~Conti and M.~Fern\'{a}ndez, \emph{Nilmanifolds with a calibrated
  {$G_2$}-structure},
  \href{https://doi.org/10.1016/j.difgeo.2011.04.030}{\emph{Differential Geom.
  Appl.} {\bfseries 29} (2011) 493}.

\bibitem{MR3739330}
L.~Bagaglini, M.~Fern\'{a}ndez and A.~Fino, \emph{Coclosed {$\rm
  G_2$}-structures inducing nilsolitons},
  \href{https://doi.org/10.1515/forum-2016-0238}{\emph{Forum Math.} {\bfseries
  30} (2018) 109}.

\bibitem{MR4626831}
G.~Bazzoni, A.~Garv\'{\i}n and V.~Mu\~{n}oz, \emph{Purely coclosed {$\rm
  G_2$}-structures on nilmanifolds},
  \href{https://doi.org/10.1002/mana.202100665}{\emph{Math. Nachr.} {\bfseries
  296} (2023) 2236}.

\bibitem{MR4789076}
G.~Bazzoni and A.~Gil-Garc\'{\i}a, \emph{Moduli spaces of (co)closed {${\rm
  G}_2$}-structures on nilmanifolds},
  \href{https://doi.org/10.1093/qmath/haae037}{\emph{Q. J. Math.} {\bfseries
  75} (2024) 987}.

\bibitem{deIaOssa:2019cci}
X.~de~Ia~Ossa, M.~Larfors, M.~Magill and E.~E. Svanes, \emph{{Superpotential of
  three dimensional $ \mathcal{N} $ = 1 heterotic supergravity}},
  \href{https://doi.org/10.1007/JHEP01(2020)195}{\emph{JHEP} {\bfseries 01}
  (2020) 195} [\href{https://arxiv.org/abs/1904.01027}{{\ttfamily
  1904.01027}}].

\bibitem{delaOssa:2021cgd}
X.~de~la Ossa, M.~Larfors and M.~Magill, \emph{{Almost contact structures on
  manifolds with a $G_2$ structure}},
  \href{https://doi.org/10.4310/ATMP.2022.v26.n1.a3}{\emph{Adv. Theor. Math.
  Phys.} {\bfseries 26} (2022) 143}
  [\href{https://arxiv.org/abs/2101.12605}{{\ttfamily 2101.12605}}].

\bibitem{delaOssa:2024dzo}
X.~de~la Ossa, M.~Larfors, M.~Magill and E.~E. Svanes, \emph{{Quantum aspects
  of heterotic $G_2$ systems}},
  \href{https://arxiv.org/abs/2412.14715}{{\ttfamily 2412.14715}}.

\bibitem{Joyce:1996i}
D.~D. Joyce, \emph{Compact riemannian 7-manifolds with holonomy $g_2$. i},
  \href{https://doi.org/10.4310/jdg/1214458109}{\emph{Journal of Differential
  Geometry} {\bfseries 43} (1996) }.

\bibitem{Joyce:1996ii}
D.~D. Joyce, \emph{Compact riemannian 7-manifolds with holonomy $g\sb 2$. ii},
  \href{https://doi.org/10.4310/jdg/1214458110}{\emph{Journal of Differential
  Geometry} {\bfseries 43} (1996) }.

\bibitem{Andriolo:2018yrz}
S.~Andriolo, G.~Shiu, H.~Triendl, T.~Van~Riet, V.~Venken and G.~Zoccarato,
  \emph{{Compact G2 holonomy spaces from SU(3) structures}},
  \href{https://doi.org/10.1007/JHEP03(2019)059}{\emph{JHEP} {\bfseries 03}
  (2019) 059} [\href{https://arxiv.org/abs/1811.00063}{{\ttfamily
  1811.00063}}].

\bibitem{Dibitetto:2018ftj}
G.~Dibitetto, G.~Lo~Monaco, A.~Passias, N.~Petri and A.~Tomasiello,
  \emph{{AdS$_3$ Solutions with Exceptional Supersymmetry}},
  \href{https://doi.org/10.1002/prop.201800060}{\emph{Fortsch. Phys.}
  {\bfseries 66} (2018) 1800060}
  [\href{https://arxiv.org/abs/1807.06602}{{\ttfamily 1807.06602}}].

\bibitem{Grana:2006kf}
M.~Grana, R.~Minasian, M.~Petrini and A.~Tomasiello, \emph{{A Scan for new N=1
  vacua on twisted tori}},
  \href{https://doi.org/10.1088/1126-6708/2007/05/031}{\emph{JHEP} {\bfseries
  05} (2007) 031} [\href{https://arxiv.org/abs/hep-th/0609124}{{\ttfamily
  hep-th/0609124}}].

\bibitem{Grana:2013ila}
M.~Gra\~na, R.~Minasian, H.~Triendl and T.~Van~Riet, \emph{{Quantization
  problem in Scherk-Schwarz compactifications}},
  \href{https://doi.org/10.1103/PhysRevD.88.085018}{\emph{Phys. Rev. D}
  {\bfseries 88} (2013) 085018}
  [\href{https://arxiv.org/abs/1305.0785}{{\ttfamily 1305.0785}}].

\bibitem{Becker:2024ayh}
K.~Becker, N.~Brady, M.~Gra\~na, M.~Morros, A.~Sengupta and Q.~You,
  \emph{{Tadpole conjecture in non-geometric backgrounds}},
  \href{https://doi.org/10.1007/JHEP10(2024)021}{\emph{JHEP} {\bfseries 10}
  (2024) 021} [\href{https://arxiv.org/abs/2407.16758}{{\ttfamily
  2407.16758}}].

\bibitem{Becker:2024ijy}
K.~Becker, M.~Rajaguru, A.~Sengupta, J.~Walcher and T.~Wrase,
  \emph{{Stabilizing massless fields with fluxes in Landau-Ginzburg models}},
  \href{https://doi.org/10.1007/JHEP08(2024)069}{\emph{JHEP} {\bfseries 08}
  (2024) 069} [\href{https://arxiv.org/abs/2406.03435}{{\ttfamily
  2406.03435}}].

\bibitem{Rajaguru:2024emw}
M.~Rajaguru, A.~Sengupta and T.~Wrase, \emph{{Fully stabilized Minkowski vacua
  in the 2$^{6}$ Landau-Ginzburg model}},
  \href{https://doi.org/10.1007/JHEP10(2024)095}{\emph{JHEP} {\bfseries 10}
  (2024) 095} [\href{https://arxiv.org/abs/2407.16756}{{\ttfamily
  2407.16756}}].

\bibitem{Duff:1984sv}
M.~J. Duff, B.~E.~W. Nilsson and C.~N. Pope, \emph{{The Criterion for Vacuum
  Stability in {Kaluza-Klein} Supergravity}},
  \href{https://doi.org/10.1016/0370-2693(84)91234-6}{\emph{Phys. Lett. B}
  {\bfseries 139} (1984) 154}.

\bibitem{Duff:1986hr}
M.~J. Duff, B.~E.~W. Nilsson and C.~N. Pope, \emph{{Kaluza-Klein
  Supergravity}},
  \href{https://doi.org/10.1016/0370-1573(86)90163-8}{\emph{Phys. Rept.}
  {\bfseries 130} (1986) 1}.

\bibitem{MR739790}
C.~S. Gordon and E.~N. Wilson, \emph{Isospectral deformations of compact
  solvmanifolds}, {\emph{J. Differential Geom.} {\bfseries 19} (1984) 241}.

\bibitem{MR4767506}
K.~Gittins, C.~Gordon, M.~Khalile, I.~Membrillo~Solis, M.~Sandoval and
  E.~Stanhope, \emph{Do the {H}odge spectra distinguish orbifolds from
  manifolds? {P}art 1},
  \href{https://doi.org/10.1307/mmj/20216126}{\emph{Michigan Math. J.}
  {\bfseries 74} (2024) 571}.

\bibitem{MR4501837}
A.~Arabia, \emph{Introduction to isospectrality}, Universitext. Springer, Cham,
  [2022] \copyright 2022,
  \href{https://doi.org/10.1007/978-3-031-17123-9}{10.1007/978-3-031-17123-9}.

\bibitem{MR2698220}
M.-P. Gong, \emph{Classification of nilpotent {L}ie algebras of dimension 7
  (over algebraically closed fields and {R})}. ProQuest LLC, Ann Arbor, MI,
  1998.

\bibitem{Miao:2025rgf}
Z.~Miao, M.~Rajaguru, G.~Tringas and T.~Wrase, \emph{{T-dualities and
  scale-separated AdS$_3$ in type I}},
  \href{https://arxiv.org/abs/2509.12801}{{\ttfamily 2509.12801}}.

\bibitem{Farakos:2025bwf}
F.~Farakos and G.~Tringas, \emph{{Integer dual dimensions in scale-separated
  AdS$_{3}$ from massive IIA}},
  \href{https://doi.org/10.1007/JHEP06(2025)130}{\emph{JHEP} {\bfseries 06}
  (2025) 130} [\href{https://arxiv.org/abs/2502.08215}{{\ttfamily
  2502.08215}}].

\bibitem{Gautason:2018gln}
F.~F. Gautason, V.~Van~Hemelryck and T.~Van~Riet, \emph{{The Tension between
  10D Supergravity and dS Uplifts}},
  \href{https://doi.org/10.1002/prop.201800091}{\emph{Fortsch. Phys.}
  {\bfseries 67} (2019) 1800091}
  [\href{https://arxiv.org/abs/1810.08518}{{\ttfamily 1810.08518}}].

\bibitem{Delgado:2025crl}
M.~Delgado, S.~Reymond and T.~Van~Riet, \emph{{Black Holes, Moduli
  Stabilisation and the Swampland}},
  \href{https://arxiv.org/abs/2504.02645}{{\ttfamily 2504.02645}}.

\bibitem{Andriot:2016rdd}
D.~Andriot, G.~Cacciapaglia, A.~Deandrea, N.~Deutschmann and D.~Tsimpis,
  \emph{{Towards Kaluza-Klein Dark Matter on Nilmanifolds}},
  \href{https://doi.org/10.1007/JHEP06(2016)169}{\emph{JHEP} {\bfseries 06}
  (2016) 169} [\href{https://arxiv.org/abs/1603.02289}{{\ttfamily
  1603.02289}}].

\bibitem{Andriot:2018tmb}
D.~Andriot and D.~Tsimpis, \emph{{Laplacian spectrum on a nilmanifold,
  truncations and effective theories}},
  \href{https://doi.org/10.1007/JHEP09(2018)096}{\emph{JHEP} {\bfseries 09}
  (2018) 096} [\href{https://arxiv.org/abs/1806.05156}{{\ttfamily
  1806.05156}}].

\bibitem{Conlon:2020wmc}
J.~P. Conlon and F.~Revello, \emph{{Moduli Stabilisation and the Holographic
  Swampland}}, \href{https://doi.org/10.31526/lhep.2020.171}{\emph{LHEP}
  {\bfseries 2020} (2020) 171}
  [\href{https://arxiv.org/abs/2006.01021}{{\ttfamily 2006.01021}}].

\bibitem{Apers:2022vfp}
F.~Apers, \emph{{Aspects of AdS flux vacua with integer conformal dimensions}},
  \href{https://doi.org/10.1007/JHEP05(2023)040}{\emph{JHEP} {\bfseries 05}
  (2023) 040} [\href{https://arxiv.org/abs/2211.04187}{{\ttfamily
  2211.04187}}].

\bibitem{Ning:2022zqx}
S.~Ning, \emph{{Holographic perspectives on models of moduli stabilization in
  M-theory}}, \href{https://doi.org/10.1007/JHEP09(2022)042}{\emph{JHEP}
  {\bfseries 09} (2022) 042}
  [\href{https://arxiv.org/abs/2206.13332}{{\ttfamily 2206.13332}}].

\bibitem{Apers:2025pon}
F.~Apers, M.~Montero and I.~Valenzuela, \emph{{Backtracking AdS flux vacua}},
  \href{https://arxiv.org/abs/2506.03314}{{\ttfamily 2506.03314}}.

\bibitem{Emelin:2022cac}
M.~Emelin, F.~Farakos and G.~Tringas, \emph{{O6-plane backreaction on
  scale-separated Type IIA AdS$_{3}$ vacua}},
  \href{https://doi.org/10.1007/JHEP07(2022)133}{\emph{JHEP} {\bfseries 07}
  (2022) 133} [\href{https://arxiv.org/abs/2202.13431}{{\ttfamily
  2202.13431}}].

\bibitem{MR2217282}
A.~V. Bolsinov, H.~R. Dullin and A.~P. Veselov, \emph{Spectra of
  {S}ol-manifolds: arithmetic and quantum monodromy},
  \href{https://doi.org/10.1007/s00220-006-1543-6}{\emph{Comm. Math. Phys.}
  {\bfseries 264} (2006) 583}.

\bibitem{NIST:DLMF}
``{\it NIST Digital Library of Mathematical Functions}.''
  \url{https://dlmf.nist.gov/}, Release 1.2.4 of 2025-03-15.

\end{thebibliography}\endgroup

\end{document}